\documentclass[lettersize,journal]{IEEEtran}

\usepackage{amsmath,amsfonts}
\usepackage{array}
\usepackage[caption=false,font=normalsize,labelfont=sf,textfont=sf]{subfig}
\usepackage{textcomp}
\usepackage{stfloats}
\usepackage{url}
\usepackage{verbatim}
\usepackage{graphicx}
\usepackage{cite}
\usepackage{afterpage}
\usepackage[T1]{fontenc}
\usepackage{longtable}
\usepackage{booktabs}
\usepackage[english]{babel}
\addto\captionsenglish{}
\usepackage[hidelinks]{hyperref}
\usepackage{makecell}
\usepackage{mathtools}
\usepackage{xstring} 

\allowdisplaybreaks[1]
\newwrite\highlightsfile
\immediate\openout\highlightsfile=highlights.txt
\AtEndDocument{\immediate\closeout\highlightsfile}

\newcommand{\HighlightItem}[1]{%
  \StrLen{#1}[\hllen]%
  \ifnum\hllen>85
    \typeout{HL WARNING: item exceeds 85 chars (\the\hllen): #1}%
  \fi
  \item #1%
  \immediate\write\highlightsfile{- #1}%
}

\begin{document}

\title{Characterisation and Quantification of Data Centre Flexibility for Power System Support}

\author{Mehmet Turker Takci\textsuperscript{1}, James Day\textsuperscript{1}, and Meysam Qadrdan\textsuperscript{1}}

\IEEEaftertitletext{%
  \vspace{-0.5em}%
  \noindent\small
  \textsuperscript{1}Cardiff University, Queen's Buildings, The Parade, Cardiff CF24 3AA, United Kingdom\par
  \vspace{0.6em}
  \noindent\small\textbf{Highlights}\par
  \vspace{0.25em}\noindent\hrule height 0.5pt \vspace{0.35em}
  \begin{itemize}
    \setlength{\itemsep}{0.2em}\setlength{\parskip}{0pt}\setlength{\parsep}{0pt}
    \HighlightItem{Integrated IT/UPS/cooling model developed for a data centre.}
    \HighlightItem{Operational optimisation model developed and costs cut by over 10\%.}
    \HighlightItem{Flexibility characterised and quantified by a duration-aware flexibility assessment.}
    \HighlightItem{A comprehensive set of results for flexibility magnitude and duration is presented.}
  \end{itemize}
  \vspace{0.35em}\noindent\hrule height 0.5pt \vspace{0.6em}
  %
  \noindent\small\textbf{Corresponding author:}\par
  \vspace{0.1em} 
  \noindent\small
  James Day\par
  \textit{E-mail address:} \texttt{DayJa1@cardiff.ac.uk}\par
  \vspace{0.6em} 
}

\maketitle

\markboth{Journal of \LaTeX\ Class Files,~Vol.~XX, No.~XX, October~2025}%
{Takci \MakeLowercase{\textit{et al.}}: Characterisation and Quantification of Data Centre Flexibility}

\begin{abstract}
The rapid growth of data centres poses an evolving challenge for power systems striving toward high variable renewable energy resources and new demands. Traditionally operated as passive electrical loads, data centres, as large loads now, have the potential to become active participants that provide flexibility to the grid. However, quantifying and utilising this flexibility have not yet been fully explored. This paper presents an integrated, whole facility optimisation model to investigate the least cost operating schedule of data centres and characterise the aggregate flexibility available from data centres to the power system. The model accounts for IT workload shifting, UPS energy storage, and cooling system. Motivated by the need to alleviate the increasing strain on power systems while leveraging their untapped flexibility potential to support decarbonisation goals, this study makes two primary contributions: (i) an operational optimisation model that integrates IT scheduling, UPS operation, and cooling dynamics to establish a cost optimal baseline operation, and (ii) a duration-aware flexibility assessment that, for any given start time and requested power deviation, computes the maximum feasible duration from this baseline while respecting all operational, thermal, and recovery constraints. This method characterises the aggregate flexibility envelope. Results reveal a clear temporal structure and a notable asymmetry in flexibility provision: upward flexibility (electricity load reduction) is driven by deferring IT workload, which allows for a secondary reduction in cooling power. In contrast, downward flexibility (electricity load increase) relies on increasing power consumption of the cooling system, supported by the TES buffer, and charging the UPS, as advancing IT workload is constrained. This framework translates abstract flexibility potential into quantified flexibility magnitude and duration capability that system operators could investigate for use in services such as reserve, frequency response, and price responsive demand.
\end{abstract}

\begin{IEEEkeywords}
Data Centre, Power system flexibility, Demand side flexibility, Load Shifting, Thermal Inertia
\end{IEEEkeywords}

\section{Introduction}

\IEEEPARstart{G}{lobal} digitalisation, propelled by artificial intelligence, is driving the rapid expansion of data centres, whose electricity demand is both substantial and spatially concentrated. This growth coincides with two other major shifts in the energy landscape: the increasing share of variable renewable generation needed for decarbonisation, and the widespread adoption of new demand-side technologies such as electric vehicles and heat pumps \cite{Centrica2020LEM, EEAACER2023Flex}. The combined effect places significant strain on modern power systems by intensifying peak loads, steepening net load ramps, and aggravating local network constraints. Consequently, there is an escalating need for power system flexibility to maintain the crucial balance between electricity supply and demand.

According to the IEA, global data centre electricity consumption was around 415 TWh in 2024, with projections suggesting this will exceed 945 TWh by 2030, raising their share of global electricity demand from approximately 1.5\% to 3\% \cite{iea_energy_ai}. This expansion is not uniform, with over 80\% of the increase expected to originate from the United States and China \cite{iea_energy_ai}. Moreover, geographical clustering intensifies the impact on local grids. For instance, data centre consumption already accounts for over 20\% of national demand in Ireland, and 25\% in Virginia, USA \cite{iea_energy_ai}. In Great Britain, annual electricity consumption by data centres is forecast by National Grid ESO to rise from 7.6 TWh in 2024 to between 30–71 TWh by 2050 \cite{neso2025future}.

\afterpage{
\clearpage 
\onecolumn 
{\small
\renewcommand{\arraystretch}{1.15}
\begin{longtable}{p{0.1\textwidth} p{0.66\textwidth} p{0.12\textwidth} p{0.06\textwidth}}
\caption{Nomenclature} \label{tb:nomenclature} \\
\toprule
\textbf{Symbol} & \textbf{Definition} & \textbf{Value/Bounds} & \textbf{Unit} \\
\midrule
\endfirsthead
\multicolumn{4}{c}%
{{\bfseries \tablename\ \thetable{} -- continued from previous page}} \\
\toprule
\textbf{Symbol} & \textbf{Definition} & \textbf{Value/Bounds} & \textbf{Unit} \\
\midrule
\endhead
\midrule
\multicolumn{4}{r}{{Continued on next page}} \\
\endfoot
\bottomrule
\endlastfoot
\midrule
\multicolumn{4}{l}{\textbf{IT Workload Modelling (Section 3.1)}} \\
\midrule
$t, s$ & Time slot an IT job originates in ($t$) and time slot the job is executed in ($s$). & [1,...,96+$D^{\text{max}}$] & - \\
$T$ & Set of time slots over a 24-hour planning horizon, representing potential arrival times for jobs. & $\{1, \dots, 96\}$ & - \\
$T^{\mathrm{ext}}$ & Extended planning horizon, allowing jobs arriving near the end of the horizon to be deferred within allowable bounds. & \{1,...,96+$D^{\text{max}}$\} & - \\
$k$ & Index for tranches of a flexible IT job. & [1,2,3,4] & - \\
$K$ & Set of tranche indices, with each tranche representing a distinct deferral class. Redefined in Section \ref{sec:scenario 3}. & $\{1, 2, 3, 4\}$ & - \\
$W_{t,k}$ & Feasible execution window of time slots for tranche k of a job originating at time t. & Varies & - \\
$\Delta t$ & Duration of a single time slot. & 0.25 & hours \\
$D_k$ & Maximum deferral duration (in time slots) for tranche k of a job, relative to its arrival time. Redefined in Section \ref{sec:scenario 3}. & $\{2, 4, 8, 12\}$ & slots \\
$P^{\mathrm{IT}}_{\mathrm{idle}}$ & Power consumption of IT equipment at idle CPU utilisation. & 166.7 &kW \\
$P^{\mathrm{IT}}_{\mathrm{max}}$ & Power consumption of IT equipment at maximum CPU utilisation. & 1000 &kW \\
$u_{\max}$ & Maximum available CPU capacity in any single time slot. & 1 & - \\
$u_t^{\mathrm{inflex}}$ & CPU utilisation required for the inflexible workload in time slot t. & [0–1] & - \\
$u_t^{\mathrm{flex,base}}$ & Base CPU utilisation required to process the flexible job originating at slot t without any deferral. & [0–1] & - \\
$R_t$ & Total computational demand (in CPU-hours) for the flexible job originating in time slot t. & Varies & CPU-hours \\
$\alpha_{t,k}$ & Fraction of the total computational demand $R_t$ that is assigned to tranche k. & [0–1] & - \\
$u(t)$ & CPU utilisation at time t, as a fraction of total capacity. & [0–1] & - \\
$u(t,k,s)$ & Decision variable for the CPU utilisation of tranche k from a job originating at t, scheduled for execution in slot s. & $\geq 0$ & - \\
$P^{\mathrm{IT}}(t)$ & Total power consumption of IT equipment at time t. & [0-1000] &kW \\
$E^{\mathrm{IT}}_{\mathrm{base}}(t)$ & Base IT energy consumption at time t. & Varies &kWh \\
$E^{\mathrm{IT}}_{\mathrm{opt}}(t)$ & Optimised IT energy consumption at time t after optimisation. & Varies &kWh \\
$P^{\mathrm{IT}}_{\mathrm{base}}(t)$ & Base IT power consumption at time t. & [0-1000] &kW \\
$P^{\mathrm{IT}}_{\mathrm{opt}}(t)$ & Optimised IT power consumption at time t after optimisation. & [0-1000] &kW \\
\midrule
\multicolumn{4}{l}{\textbf{UPS-ESS Modelling (Section 3.2)}} \\
\midrule
$E^{\mathrm{UPS}}_{\mathrm{base}}$ & Base rated energy capacity of the UPS battery system. & 600 &kWh \\
$SoC_{\min}$ & Minimum state of charge, expressed as a fraction of $E_{\mathrm{UPS}}^{\mathrm{base}}$. & 0.5 & - \\
$SoC_{\max}$ & Maximum state of charge, expressed as a fraction of $E_{\mathrm{UPS}}^{\mathrm{base}}$. & 1.0 & - \\
$P^{\mathrm{UPS}}_{\mathrm{ch,min}}$ & Minimum allowable charging power of the UPS. & 40 &kW \\
$P^{\mathrm{UPS}}_{\mathrm{ch,max}}$ & Maximum allowable charging power of the UPS. & 270 &kW \\
$P^{\mathrm{UPS}}_{\mathrm{disch,min}}$ & Minimum allowable discharging power of the UPS. & 100 &kW \\
$P^{\mathrm{UPS}}_{\mathrm{disch,max}}$ & Maximum allowable discharging power of the UPS. & 2700 &kW \\
$\eta^{\mathrm{UPS}}_{\mathrm{ch}}$ & Charging efficiency of the UPS battery system. & 0.82 & - \\
$\eta^{\mathrm{UPS}}_{\mathrm{disch}}$ & Discharging efficiency of the UPS battery system. & 0.92 & - \\
$E^{\mathrm{UPS}}(t)$ & Energy stored in the UPS battery system at the end of time slot t. & $[300, 600]$ &kWh \\
$P^{\mathrm{UPS}}_{\mathrm{ch}}(t)$ & Charging power drawn by the UPS from the grid during time slot t. & $\geq 0$ &kW \\
$P^{\mathrm{UPS}}_{\mathrm{disch}}(t)$ & Discharging power supplied by the UPS during time slot t. & $\geq 0$ &kW \\
$P^{\mathrm{UPS}}_{\mathrm{net}}(t)$ & Net power exchange of the UPS at time t, defined as $P^{\mathrm{UPS}}_{\mathrm{ch}}(t) - P^{\mathrm{UPS}}_{\mathrm{disch}}(t)$. & Varies &kW \\
$z^{\mathrm{UPS}}_{\mathrm{ch}}(t)$ & Binary variable indicating charging status in time slot t. & $\{0, 1\}$ & - \\
$z^{\mathrm{UPS}}_{\mathrm{disch}}(t)$ & Binary variable indicating discharging status in time slot t. & $\{0, 1\}$ & - \\
\midrule
\multicolumn{4}{l}{\textbf{Cooling System Modelling (Section 3.3)}} \\
\midrule
$\dot{m}_a$ & Constant mass flow rate of air circulated by the cooling unit. & 100 & kg/s \\
$c_{pa}$ & Specific heat capacity of air at constant pressure. & 1.005 & kJ/(kg$\cdot$K) \\
$\rho_a$ & Density of air. & 1.16 & kg/m$^3$ \\
$C_{IT}$ & Total heat capacity of the IT components. & $1.788 \times 10^4$ & kJ/K \\
$C_R$ & Total heat capacity of the server racks and the air within them. & $1.802 \times 10^4$ & kJ/K \\
$C_{CA}$ & Total heat capacity of the air in the cold aisle. & $2.33 \times 10^3$ & kJ/K \\
$C_{HA}$ & Total heat capacity of the air in the hot aisle. & $1.17 \times 10^3$ & kJ/K \\
$G_{cv}$ & Convective heat conductance coefficient between IT components and rack air. & 109 &kW/K \\
$G_{cd}$ & Total heat transfer coefficient from the cold aisle to the outside environment. & 4.484 &kW/K \\
$\kappa$ & Correction factor for the effectiveness of the air mass flow through racks. & 0.766 & - \\
$COP_{\mathrm{chiller}}$ & Coefficient of Performance of the chiller. & 5 & - \\
$P^{\mathrm{Chiller}}_{\mathrm{max}}$ & Maximum electrical power consumption capacity of the chiller. & 400 &kW \\
$E^{\mathrm{TES}}_{\mathrm{max}}$ & Maximum cooling energy storage capacity of the TES tank. & 1000 &kWh \\
$Q^{\mathrm{TES}\text{-}\mathrm{CRAC}}_{\mathrm{max}}$ & Maximum thermal charging rate of the TES tank. & 300 &kW \\
$Q^{\mathrm{Chil}\text{-}\mathrm{TES}}_{\mathrm{max}}$ & Maximum thermal discharging rate of the TES tank. & 300 &kW \\
$\eta^{\mathrm{TES}}_{\mathrm{ch}}$ & Charging efficiency of the TES tank. & 0.9 & - \\
$\eta^{\mathrm{TES}}_{\mathrm{dis}}$ & Discharging efficiency of the TES tank. & 0.9 & - \\
$T_{\mathrm{out}}$ & Outside ambient air temperature. & 22 & $^{\circ}$C \\
$Q^{\mathrm{IT}}(t)$ & Heat generated by ITE, assumed equal to $P^{\mathrm{IT}}(t)$. & [0-1000] &kW \\
$Q_{\mathrm{cool}}(t)$ & Total cooling power provided to the DC from the CRAC and TES. & $\geq 0$ &kW \\
$Q^{\mathrm{Chil}\text{-}\mathrm{CRAC}}(t)$ & Cooling power provided directly from the chiller to the CRAC system. & Varies &kW \\
$Q^{\mathrm{Chil}\text{-}\mathrm{TES}}(t)$ & Cooling power from the chiller used to charge the Thermal Energy Storage (TES) tank. & $[0, 300]$ &kW \\
$Q^{\mathrm{TES}\text{-}\mathrm{CRAC}}(t)$ & Cooling power discharged by the TES tank to the CRAC system. & $[0, 300]$ &kW \\
$Q_{CA}(t)$ & Rate of change of thermal energy in the air mass of the cold aisle. & Varies &kW \\
$Q_{HA}(t)$ & Rate of change of thermal energy in the air mass of the hot aisle. & Varies &kW \\
$Q_{R}(t)$ & Rate of change of thermal energy in the mass of the server racks and the air within them. & Varies &kW \\
$Q_{ITm}(t)$ & Rate of change of thermal energy in the thermal mass of the IT equipment itself. & Varies &kW \\
$Q_{\mathrm{out}}(t)$ & Thermal energy transferred into the DC from the outside ambient environment. & Varies &kW \\
$P^{\mathrm{Chil}\text{-}\mathrm{CRAC}}(t)$ & Electrical power consumed by the chiller for cooling sent directly to the CRAC. & $\geq 0$ &kW \\
$P^{\mathrm{Chil}\text{-}\mathrm{TES}}(t)$ & Electrical power consumed by the chiller to charge the TES tank. & $\geq 0$ &kW \\
$E^{\mathrm{TES}}(t)$ & Cooling energy stored in the TES tank at time t. & $[0, 1000]$ &kWh \\
$T_{Ain}(t)$ & Air temperature at the inlet of the cold aisle. & $[14, 30]$ & $^{\circ}$C \\
$T_{CA}(t)$ & Air temperature in the cold aisle. & $[18, 22.5]$ & $^{\circ}$C \\
$T_{HA}(t)$ & Air temperature in the hot aisle. & $[18, 40]$ & $^{\circ}$C \\
$T_{R}(t)$ & Air temperature within the server racks. & $[18, 40]$ & $^{\circ}$C \\
$T_{IT}(t)$ & Temperature of the IT components. & $[18, 60]$ & $^{\circ}$C \\
$z^{\mathrm{Chil}\text{-}\mathrm{TES}}(t)$ & Binary variable indicating TES charging status (1 if charging, 0 otherwise). & $\{0, 1\}$ & - \\
$z^{\mathrm{TES}\text{-}\mathrm{CRAC}}(t)$ & Binary variable indicating TES discharging status (1 if discharging, 0 otherwise). & $\{0, 1\}$ & - \\
\midrule
\multicolumn{4}{l}{\textbf{Case Studies (Section 4)}} \\
\midrule
$\pi(t)$ & Day-ahead spot electricity price at time t. & Varies & GBP/MWh \\
$P_{\mathrm{tol}}$ & Power tolerance for flexibility provision. & 0.1 &kW \\
$P^{\mathrm{Grid}}_{\mathrm{IT}}(t)$ & Power drawn from grid for IT load. & $\geq 0$ &kW \\
$P^{\mathrm{Grid}}_{\mathrm{OD}}$ & Power drawn from grid for auxiliary devices. & $53.095$ &kW \\
$P^{\mathrm{Grid}}_{\mathrm{base}}(t)$ & Baseline power drawn from grid at time t. & Varies &kW \\
$P^{\mathrm{Grid}}(t)$ & Total power drawn from grid at time t. & Varies &kW \\
$\Delta P$ & Flexibility magnitude (power deviation from baseline). & Varies &kW \\
$\tau$ & Maximum duration of flexibility provision. & [0–23:45] & hours \\
$t_0$ & Start time of flexibility provision. & [0–23:45] & hours \\
\end{longtable}
\twocolumn
}
} 
To manage the volatility introduced by these trends, power systems require a substantial increase in flexibility services. Traditionally provided by dispatchable fossil fuel generators, this solution is becoming obsolete under decarbonisation mandates. The JRC \cite{koolen2023flexibility} projects that the European Union’s flexibility requirement will rise to 24\% of total electricity demand in TWh by 2030 and to 30\% by 2050. Similarly, Great Britain’s Clean Flexibility Roadmap targets an expansion of flexibility capacity from 25.2 GW to between 54–66 GW by 2030 \cite{clean_flexibility_roadmap}. Demand-side flexibility, where consumers actively modulate their electricity use to support the grid, presents a key solution.

While their growth contributes to the challenge, data centres are equipped to provide flexibility. Their shiftable IT workloads, uninterruptible power supply (UPS) systems with battery storage, and inherent thermal inertia as well as cold storage, can be leveraged to modulate power consumption without disrupting core services. Each of these internal assets possesses a distinct flexibility profile in terms of magnitude, duration, ramp rate, and recovery time \cite{zhang2025unlocking}. For example, UPS batteries can offer a near instantaneous response, whereas thermal systems are characterised by lower ramp rates, providing a more gradual adjustment \cite{zhang2025unlocking, alapera2018data}. Data centres therefore present a dichotomy, they are both a growing strain on the power system and a potentially crucial asset for grid flexibility.

In the context of power systems, flexibility is defined as the capability of maintaining the balance between generation and demand across all time horizons, by adjusting either energy production or consumption in response to sudden variations \cite{Centrica2020LEM, EEAACER2023Flex, neso2025future, Nosair2015TSTE}. Upward flexibility is required when demand exceeds generation and involves a DC reducing its power consumption from the grid. Conversely, downward flexibility is needed when generation surpasses demand, and is provided by a DC increasing its electricity consumption to absorb the surplus energy \cite{Centrica2020LEM, NG2017SNAPS}. In this study, a data centre's flexibility is quantified by its capacity for power consumption deviation relative to a baseline, and the maximum duration for which this deviation can be sustained.

This paper addresses these challenges through two main contributions. First, we develop a least cost optimisation model for scheduling IT workload, UPS dispatch, and cooling system operation, considering day-ahead electricity prices. Second, we introduce a methodology to quantify the dynamic aggregate flexibility capacity and duration a data centre can offer to the power grid. Our analysis reveals the asymmetric nature of upward and downward flexibility, detailing the distinct roles of each internal asset. By providing a method to develop an in-depth understanding of data centres' flexibility potential, this work sheds light on how data centres can be more efficiently integrated into our power system.

The remainder of this paper is structured as follows. Section 2 reviews the existing literature on data centre flexibility. Section 3 details the mathematical modelling of the key data centre components. Section 4 outlines the case studies used to test the model. Section 5 presents and discusses the results, and Section 6 concludes the paper. 

\subsection{Data Centre Architecture}
\begin{figure}[!t]
    \centering
    \includegraphics[width=\columnwidth]{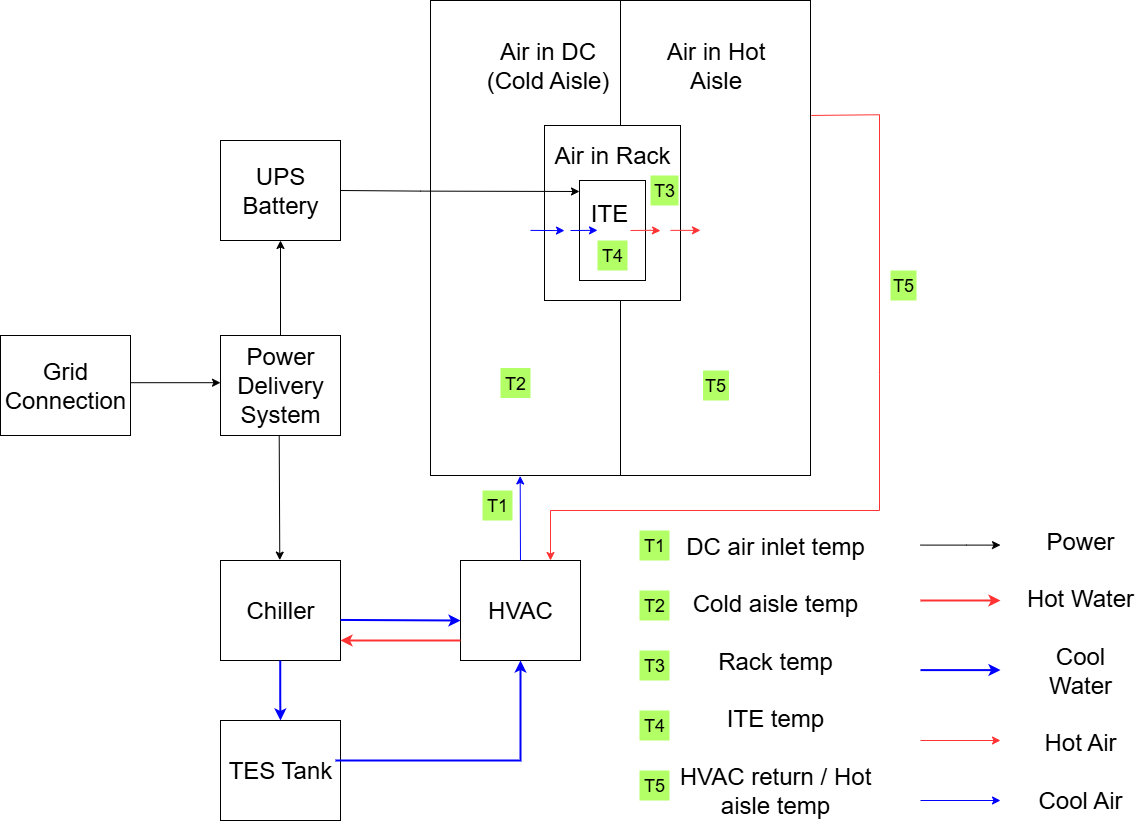}
    \caption{Data Centre Layout}
    \label{fig:dc_layout}
\end{figure}

DCs are integrated facilities housing IT equipment, such as servers and network devices, which are supported by critical power and cooling infrastructure. As shown in Figure~\ref{fig:dc_layout}, electrical continuity is maintained by uninterruptible power supplies (UPS) with associated battery systems, while cooling systems, including chillers, Thermal Energy Storage (TES) tanks and  Computer Room Air Conditioning (CRAC) units, manage the substantial heat generated by the servers. Thermal efficiency is often enhanced through optimised airflow strategies like the hot-aisle/cold-aisle arrangement.

Each of these core subsystems presents an opportunity for providing demand-side flexibility to the power grid. The strategic scheduling of IT workloads, the dispatch of UPS battery storage, and the leveraging of the TES tank and the facility's inherent thermal inertia allow a DC to modulate its power consumption. The DC assets are highly interdependent, requiring an integrated approach that considers all of them simultaneously. Consequently, DCs can transition from being passive electrical loads to active assets that enhance the resilience and flexibility of the power system.

\subsection{IT Workload Background}
The fundamental role of a DC is to execute IT workloads, which encompass all computational and data processing tasks. These workloads can be broadly classified based on their time sensitivity, as defined by service-level agreements (SLAs) \cite{Liu2024Access,Cao2022ApEn}. Inflexible workloads, also known as interactive workloads, are latency-critical and require immediate execution to ensure quality of service for user-facing applications like real-time streaming or transaction processing.

In contrast, flexible IT workloads, often referred to as batch workloads, can tolerate execution delays within predefined time windows without impacting service quality. These typically non-user-facing tasks, such as periodic data analytics, backups, or machine learning model training, can be deferred or rescheduled. This ability to shift the timing of their execution is the primary source of IT-based flexibility, allowing the DC to adjust its computational power draw in response to external signals like electricity prices or grid operational needs \cite{Cao2022ApEn}.

\section{Literature Review}

There is extensive literature on modelling the flexibility of assets within a DC. A significant proportion of studies focus on IT workloads, with some modelling UPS batteries, the cooling system and other DC components. In this review, the flexibility modelling approaches found in the literature for each DC asset are presented. In the final section, studies that attempt to develop an integrated model are discussed.  

IT workload is the most fundamental component of a DC and is the focus for many flexibility studies. One study develops a machine learning method to predict DC energy consumption and server temperature \cite{misaghian2022assessment}. These data are then used to optimise the scheduling of IT workloads to minimise fuel costs and power utilisation at times of high grid carbon intensity. On average, the implementation achieved a 6.5\% load reduction for a winter day during periods of high CO$_2$ intensity. Radovanović et al. \cite{radovanovic2022carbon}, performed predictive modelling of grid carbon intensity, trained day-ahead demand prediction models and used them to optimise IT workload for minimum carbon emissions.
Cao et al. \cite{cao2022data} propose and validate a data-driven methodology using real-world Alibaba trace data to assess the power flexibility potential of Internet Data Centres (IDCs) by rescheduling periodic IT workloads. Their findings quantify significant flexibility, particularly in early morning hours, revealing potential upward/downward load shifts of 400kW and 325kW, respectively—corresponding to approximately 50\% and 40.6\% of the facility's peak demand—by optimising job schedules based on power system needs and renewable energy availability. In this study, IT workloads are defined as either flexible or inflexible, with flexibility here referring to flexibility in time.

UPS batteries are currently providing power flexibility to the grid in numerous DCs around the world and have been shown to be successful \cite{alapera2019fast,roach2022microsoft,Anghel_2023}. Lithium-ion batteries, which are increasingly the dominant form used for UPS, are capable of rapidly providing or consuming power when needed \cite{alapera2018data}. These characteristics make them well suited to frequency regulation, a service many power providers require. Results from \cite{alapera2018data} found UPS batteries could be used within their normal operating range to provide the required services to the grid and in doing so generate a reasonable revenue stream. Furthermore, the battery lifetime would not be affected, and the approach does not require additional cost.

Cooling makes up a large proportion, as high as 40\%, of DC energy consumption \cite{zhao2021synthetic}.  The American Society of Heating, Refrigerating and Air-Conditioning Engineers (ASHRAE) recommends maintaining an indoor temperature of $18^\circ\mathrm{C}$ to $27^\circ\mathrm{C}$ to ensure optimal DC performance \cite{tc2016data}. Exploiting the thermal inertia provided by this temperature range has been shown to enable power consumption flexibility in DCs \cite{takci2025data}\cite{ref22}. Thermal Energy Storage (TES) systems can enhance cooling flexibility by decoupling the timing of heat removal from the cooling generation process.
Du et al. \cite{du2025new} propose and evaluate a framework to enhance the energy flexibility of district heating (DH) systems integrated with DC waste heat recovery. The study utilises dual short-term TES tanks to enable simultaneous peak shaving (demand-driven) and load shifting (price-driven) within the hybrid system. Simulation results demonstrated that this dual TES approach significantly improved flexibility, achieving up to a 10\% peak demand reduction and a 2.1\% load shift, leading to a 3.2\% operational cost saving in the case study. A `synergistic control strategy for data centre frequency regulation which uses both IT and cooling systems' has also been proposed \cite{fu2020assessments}. Results found that a revenue saving of 4\% could be made by implementing the proposed strategy. Once depleted, the thermal inertia of the DC atmosphere / TES must be recovered to baseline levels. The recovery time taken to do so will affect how the flexibility of the asset can be utilised.

While individual assets offer flexibility, the flexibility characteristics change when these assets are integrated into one system. Maximising the overall potential requires coordinating multiple resources and evaluating their diverse characteristics, interactions, and constraints. One study proposes a co-optimisation of the IT workload and cooling system, recognising the strong operational link between them \cite{fu2020assessments}. The work by S Xiang et al. \cite{xiang2023modeling} identifies that data centre optimisation models often ignore IT equipment constraints like start-stop conditions and ramp rates, which can cause excessive wear. The authors propose a model integrating these constraints with facility thermal dynamics, solved using a Deep Deterministic Policy Gradient (DDPG) algorithm. This approach was then validated for practicality and efficiency using a large-scale simulation of a 100,000-device data centre.
H. Xu et al. \cite{xu2023data} propose an optimisation which integrates IT servers, cooling infrastructure, energy storage, generators and renewable energy, and aims to maximise power demand flexibility. The objective function is the cumulative change in energy taken from the grid before and after the optimisation was applied. 
Other studies use a similar approach, but instead minimise the difference between DC power consumption and the desired power consumption of the power supply operator \cite{cioara2018optimized}. 

Few studies propose an integrated DC model that characterises the flexibility of DC assets and evaluates their potential for providing this flexibility to the power system. Furthermore, few studies analyse the complex interplay between DC assets and how an integrated model can improve the flexibility potential. Many studies take a DC centric view, where optimising costs is the main objective. However, as data centres scale up to the gigawatt level, a power system centric view that investigates their potential benefits to the power system is required.

Addressing these gaps, in this study, a flexibility analysis is performed to determine, for a given start time and flexibility magnitude, the maximum duration for which this flexibility can be provided. This analysis is run for a large range of flexibility magnitudes and start times to establish a heatmap of flexibility potential. For each flexibility calculation, the contribution that each DC asset makes is investigated. This analysis determines the contribution of each asset to the overall flexibility and reveals how the characteristics of each asset are exploited to maximise the flexibility potential. From a power-system-centric view, the magnitude and duration of flexibility is what quantifies its utility. Thus, by quantifying these values, the benefit the model data centre can provide to the power system is determined. 

\section{Methodology}
This section details the mathematical framework developed to quantify the demand-side flexibility of a data centre by co-optimising its primary subsystems. The conceptual model of the data centre, illustrated in Figure~\ref{fig:dc_layout}, integrates three dynamically coupled components: the IT systems, which process both flexible and inflexible computational workloads; the power infrastructure, featuring an Uninterruptible Power Supply which operates as an Energy Storage System (UPS-ESS); and the cooling infrastructure, which includes a chiller coupled with a Thermal Energy Storage (TES) tank and the CRAC system. 

Our approach is centred on a comprehensive optimisation model that captures the operational dynamics and physical constraints of each subsystem. Flexibility is harnessed from three key sources: (1) shifting  IT workloads in time, (2) strategically charging and discharging the UPS-ESS, and (3) strategically managing the cooling system by utilising the TES and the inherent thermal inertia of the facility. By integrating these components into a unified framework, the data centre's total capacity to modulate its electricity consumption is evaluated, thereby quantifying its potential to provide valuable grid services.

\label{sec: methodology}
\subsection{IT Workload Methodology}
 
In the literature, there is no consensus on IT workload characteristics and utilisation levels, as they vary significantly depending on the data centre size, type (e.g., enterprise, colocation, hyperscale) and workload nature (e.g., banking, AI). Consequently, even publicly available real-world datasets can differ substantially across facilities. Chris Zaloumis \cite{Zaloumis2022IBM} indicates that average CPU utilisations vary between 12\% and 18\% of total capacity, while Ankur Ghia's findings indicate a daily average between 20\% and 30\%, noting that best practices can elevate this to 70\%--80\% \cite{Ghia2011McKinsey}. 
 
 A study by Google \cite{6779441} classified data centre workloads into four categories based on their sensitivity to delay, labelled from ``0'' (least sensitive) to ``3'' (most sensitive). If categories 2 and 3 are considered inflexible, they account for approximately 30\% of the total workloads, suggesting that up to 70\% of the workloads could be flexible. Another study by Cao et al. \cite{Cao2022ApEn}, conducted using Alibaba’s cluster trace, found that inflexible workloads represent 60\% of all jobs but contribute only 30\% of total energy consumption. In contrast, flexible workloads constitute 40\% of jobs while consuming 70\% of the energy. Furthermore, studies such as \cite{Misaghian2023TIA,Xu2023SPIE,Zhou2024IJEPES,Fu2020ApEn,Liu2012PER} reveal that there is no universally accepted ratio, and the estimated shares of flexible and inflexible IT workloads differ significantly across studies for a typical 24-hour period. 

Based on the review of the literature, a dataset of reported workload ratios and graphs over a 24-hour time horizon was compiled. From this dataset, the minimum and maximum values were extracted, and the average ratio was computed, separately for flexible, inflexible, and total workloads relative to total CPU capacity. Drawing upon these values, the workload ratio assumptions adopted in this study were established and are presented under the ``Proposed'' category in Table~\ref{tab:workload-rates}.

\begin{table}[!t]
\centering
\caption{IT Workload Rates}\label{tab:workload-rates}
\begin{scriptsize}
\setlength{\tabcolsep}{3pt}
\resizebox{\columnwidth}{!}{%
\begin{tabular}{lrrrrrrrrrrrrr}
\toprule
{\textbf{Time}} & \multicolumn{4}{c}{\textbf{Flexible Workload Ratio (\%)}} & \multicolumn{4}{c}{\textbf{Inflexible Workload Ratio (\%)}} & \multicolumn{4}{c}{\textbf{Total Workload Ratio (\%)}} \\
\cmidrule(lr){2-5} \cmidrule(lr){6-9} \cmidrule(lr){10-13}
& Min & Max & Average & Proposed & Min & Max & Average & Proposed & Min & Max & Average & Proposed \\
\midrule
00--01 & 17 & 75 & 37 & 40 & 20 & 39 & 32 & 28 & 55 & 80 & 68 & 68 \\
01--02 & 13 & 72 & 36 & 31 & 25 & 38 & 32 & 25 & 46 & 66 & 56 & 56 \\
02--03 & 18 & 70 & 38 & 33 & 25 & 37 & 32 & 17 & 43 & 50 & 47 & 50 \\
03--04 & 15 & 68 & 37 & 23 & 19 & 38 & 26 & 16 & 34 & 37 & 36 & 39 \\
04--05 & 17 & 64 & 36 & 27 & 8 & 36 & 21 & 8 & 25 & 37 & 31 & 35 \\
05--06 & 15 & 59 & 32 & 27 & 2 & 39 & 21 & 6 & 22 & 36 & 29 & 33 \\
06--07 & 14 & 51 & 29 & 18 & 13 & 40 & 25 & 12 & 20 & 36 & 28 & 30 \\
07--08 & 24 & 42 & 31 & 24 & 18 & 39 & 27 & 20 & 24 & 52 & 38 & 44 \\
08--09 & 23 & 35 & 30 & 24 & 30 & 53 & 41 & 24 & 31 & 65 & 48 & 48 \\
09--10 & 27 & 49 & 35 & 28 & 36 & 65 & 49 & 34 & 38 & 85 & 62 & 62 \\
10--11 & 23 & 50 & 35 & 19 & 37 & 62 & 51 & 37 & 45 & 87 & 66 & 56 \\
11--12 & 19 & 50 & 35 & 18 & 42 & 70 & 56 & 42 & 50 & 92 & 71 & 60 \\
12--13 & 17 & 49 & 33 & 20 & 40 & 51 & 47 & 40 & 52 & 89 & 71 & 60 \\
13--14 & 16 & 49 & 33 & 24 & 36 & 60 & 49 & 36 & 54 & 85 & 70 & 60 \\
14--15 & 16 & 52 & 35 & 27 & 35 & 68 & 52 & 35 & 55 & 87 & 71 & 62 \\
15--16 & 17 & 49 & 35 & 27 & 33 & 70 & 53 & 33 & 57 & 82 & 70 & 60 \\
16--17 & 26 & 43 & 37 & 20 & 35 & 62 & 52 & 40 & 63 & 78 & 71 & 60 \\
17--18 & 37 & 39 & 38 & 40 & 20 & 60 & 39 & 27 & 67 & 75 & 71 & 67 \\
18--19 & 32 & 49 & 39 & 45 & 10 & 63 & 37 & 26 & 70 & 73 & 72 & 71 \\
19--20 & 23 & 54 & 35 & 45 & 9 & 67 & 38 & 26 & 62 & 77 & 70 & 71 \\
20--21 & 22 & 58 & 36 & 47 & 8 & 62 & 37 & 25 & 63 & 83 & 73 & 72 \\
21--22 & 21 & 61 & 38 & 41 & 7 & 61 & 37 & 29 & 63 & 87 & 75 & 70 \\
22--23 & 17 & 64 & 38 & 40 & 6 & 55 & 35 & 30 & 61 & 89 & 75 & 70 \\
23--24 & 16 & 69 & 40 & 42 & 2 & 45 & 29 & 21 & 56 & 80 & 68 & 63 \\
\bottomrule
\end{tabular}%
}
\end{scriptsize}
\end{table}

\begin{figure}[!t]
\centering
\includegraphics[width=\columnwidth]{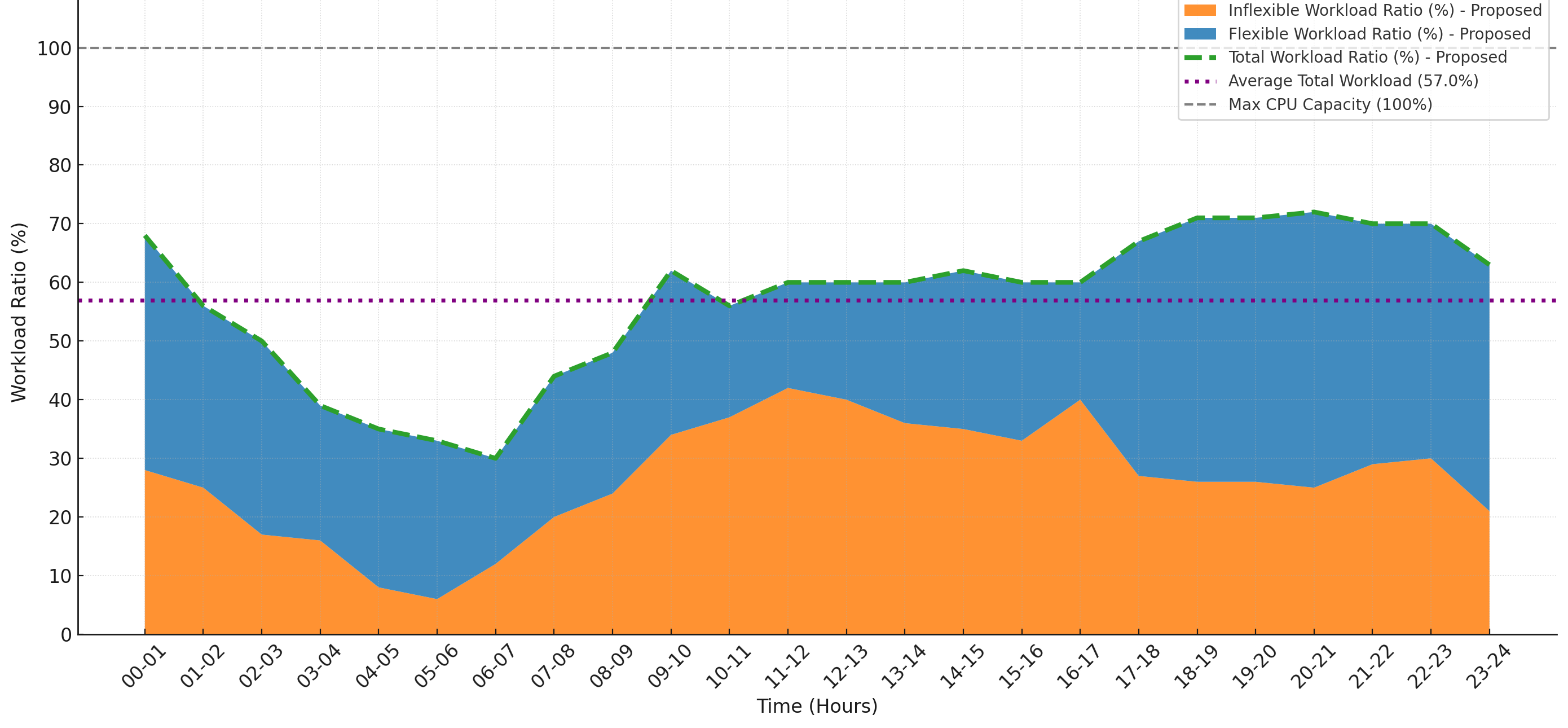}
\caption{Stacked Proposed Workload Ratios Over 24 Hours}
\label{fig:workload_ratios}
\end{figure}

Categorising the deferral flexibility of IT workloads, defined as the proportion of workloads that can be shifted and the maximum duration by which they can be postponed, is essential for developing effective load-shifting strategies. Drawing on insights from various studies \cite{Centrica2020LEM, Cao2022ApEn}, a representative set of deferral windows applicable to flexible IT workloads over a 24-hour period is defined and summarised in Table~\ref{tab:shiftable}.

\begin{table}[!t]
\centering
\caption{Hourly Distribution of Shiftable Workload Under Different Deferral Windows (\%)}
\label{tab:shiftable}
\setlength{\tabcolsep}{2pt} 
\begin{scriptsize}
\resizebox{\columnwidth}{!}{%
\begin{tabular}{l|*{24}{c}}
\toprule
\textbf{Deferral} & \multicolumn{24}{c}{\textbf{Time}} \\
\cmidrule(l){2-25}
& 00 & 01 & 02 & 03 & 04 & 05 & 06 & 07 & 08 & 09 & 10 & 11 & 12 & 13 & 14 & 15 & 16 & 17 & 18 & 19 & 20 & 21 & 22 & 23 \\
\midrule
0 $\leq$ 30 min & 25 & 25 & 35 & 25 & 25 & 25 & 23 & 20 & 25 & 25 & 35 & 32 & 45 & 50 & 40 & 45 & 50 & 50 & 10 & 15 & 18 & 22 & 20 & 15 \\
0 $\leq$ 60 min & 25 & 25 & 17 & 15 & 20 & 15 & 28 & 20 & 15 & 17 & 22 & 20 & 25 & 20 & 20 & 25 & 15 & 20 & 20 & 20 & 12 & 18 & 16 & 20 \\
0 $\leq$ 2 h & 20 & 15 & 18 & 20 & 15 & 15 & 15 & 15 & 13 & 16 & 13 & 15 & 15 & 15 & 25 & 15 & 20 & 15 & 30 & 25 & 25 & 25 & 24 & 20 \\
0 $\leq$ 3 h & 30 & 35 & 30 & 40 & 40 & 45 & 34 & 45 & 47 & 42 & 30 & 33 & 15 & 15 & 15 & 15 & 15 & 15 & 40 & 40 & 45 & 35 & 40 & 45 \\
\bottomrule
\end{tabular}%
}
\end{scriptsize}
\end{table}

Table~\ref{tab:shiftable} presents the hourly distribution of flexible workloads across four maximum deferral windows ($\leq$ 30 minutes, $\leq$ 60 minutes, $\leq$ 2 hours, and $\leq$ 3 hours). The percentages represent the relative distribution within the flexible workload. For instance, during the 12:00--13:00 interval, 45\% of the flexible workloads can be deferred up to 30 minutes, 25\% for up to 60 minutes, 15\% for up to 2 hours, and the remaining 15\% for up to 3 hours. These categorized hourly workload profiles are shown in Figure \ref{fig:workload_ratios}  
and utilised as modelling inputs, capturing the time-dependent and duration-sensitive characteristics of data centre flexibility in this study.

\subsubsection{IT Power Consumption Model}

 In the literature, various approaches exist for modelling the power consumption of IT equipment, particularly servers. Given that server power consumption often constitutes the dominant portion of the total IT power draw, other components are frequently disregarded, and server power consumption is typically taken as a proxy for the overall IT power consumption. Numerous linear and non-linear models have been developed to represent server power consumption. A fundamental and widely accepted principle is that server power consumption is strongly correlated with its CPU utilisation. Equation~\ref{eq:power_it} was used in this study to estimate the electric power consumption associated with the IT workload \cite{dayarathna2015data,v2015analysis}.
\begin{equation}
P^{\mathrm{IT}}(t) = P^{\mathrm{IT}}_{\mathrm{idle}} + (P^{\mathrm{IT}}_{\mathrm{max}} - P^{\mathrm{IT}}_{\mathrm{idle}}) \times u(t)^{1.32}
\label{eq:power_it}
\end{equation}

Therefore, the flexibility provided through IT workload shifting fundamentally depends on the rescheduling of CPU utilisation over time.

\begin{figure}[!t]
 \centering
 \includegraphics[width=\columnwidth]{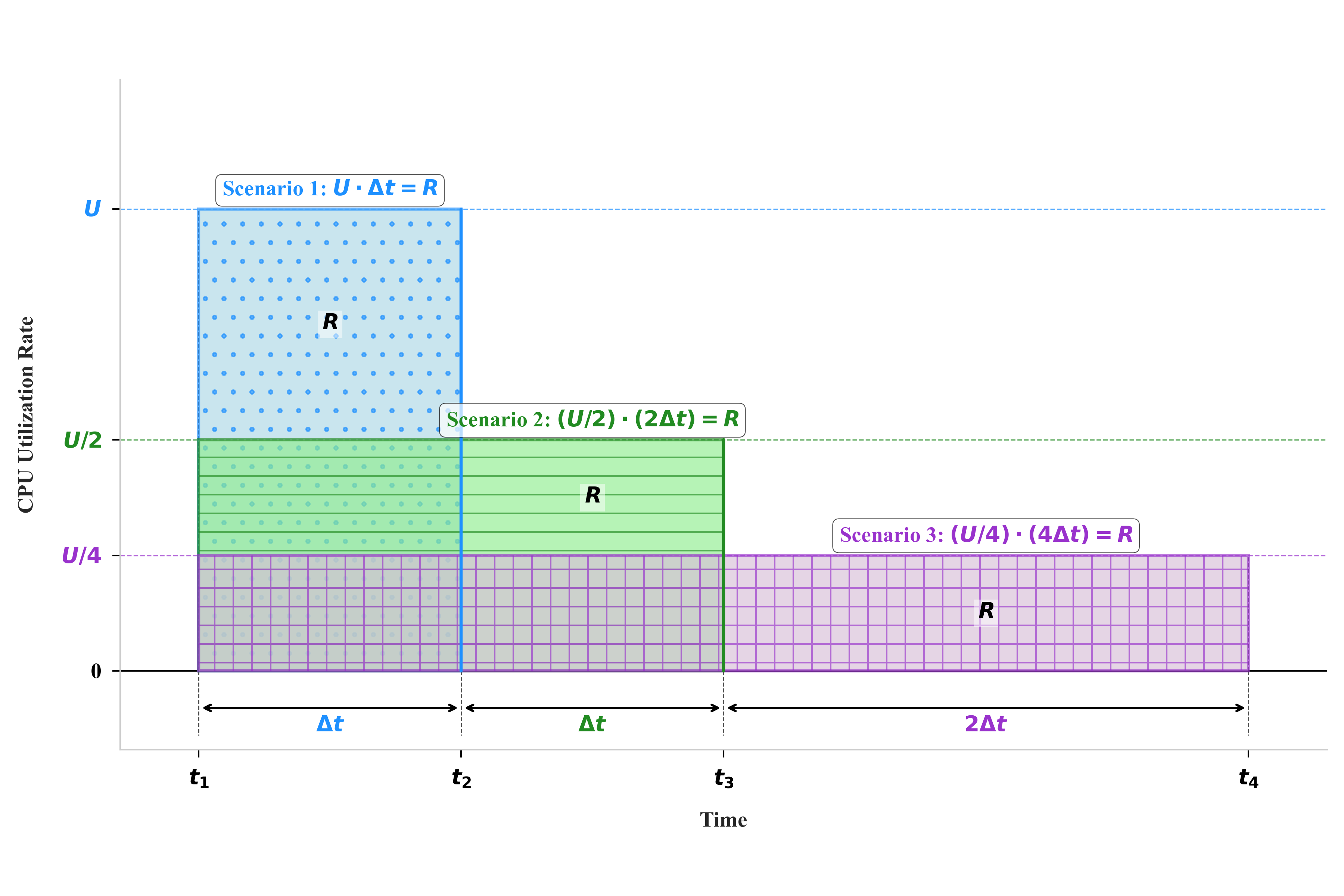}
 \caption{Constant Computational Demand ($R$) with Varying CPU Utilisation and Duration}
 \label{fig:constantR}
\end{figure}

The core principle of workload shifting employed in this article relies on the concept of constant computational demand for each job. This demand, denoted as $R$, is defined by the required CPU resources and the duration for which they are needed. Execution can be at a lower or higher CPU utilisation for a longer or shorter duration respectively. In all cases, the IT job must be fully executed within the maximum delay tolerance defined by its tranche. The crucial relationship of constant computational demand for a given IT job is mathematically represented in Equation~\ref{eq:comp_demand} and conceptually illustrated in Figure~\ref{fig:constantR}.
\begin{equation}
R = u(t) \times \Delta T
\label{eq:comp_demand}
\end{equation}

As depicted in Figure~\ref{fig:constantR}, a job with a constant computational demand $R$, can be executed over varying time frames $\Delta T$, by adjusting the allocated CPU utilisation $u(t)$. Decreasing CPU utilisation extends the job’s duration, while increasing it shortens the duration, all while maintaining the same total computational demand ($R$). Consequently, the power consumption profile corresponding to the CPU utilisation changes accordingly. By strategically increasing or decreasing CPU utilisation in response to system requirements or grid signals, the associated power consumption can be modulated, enabling the provision of upward or downward flexibility, respectively.

\subsubsection{IT Workload Formulation} \label{sec:Quantifying Maximum Temporal Flexibility from IT Workload Shifting}
The IT workload formulation presented in this paper accounts for both flexible and inflexible job types and incorporates the power consumption model formulated in Equation~\ref{eq:power_it}. The formulation is defined over a 24 hour time horizon, which is discretized into 15-minute time slots. A 3 hour extension period is appended to this horizon, allowing the deferral of IT workloads from the final three hours of the original 24 hour window.

The model conceptualises all servers of the entire DC as a single representative server, treating all IT workloads as if processed by a single computational unit. This simplification enables a tractable yet representative analysis of IT workload deferral at the system level.

The flexible IT workloads are decomposed into tranches as shown in Table \ref{tab:shiftable}, with each tranche representing the portion of the jobs that can be shifted up to a specified maximum delay duration. The primary decision variable is $u(t,k,s)$, which denotes the CPU utilisation of tranche $k$, of a job originating at slot $t \in T$, which is executed at time slot $s \in W_{t,k}$. IT workload shifting and execution is achieved while satisfying constraints \eqref{eq:power_optimised}--\eqref{eq:cpu_capacity_new}.

The allocated CPU utilisation for any tranche must be non-negative and cannot exceed the maximum capacity, as shown in (\ref{eq:util_bounds}). The base IT power and energy consumption (i.e., without optimisation) are calculated as shown in (\ref{eq:power_base}) and (\ref{eq:energy_base}), respectively. Without load shifting, the workload in the main 24 hour period consists of both inflexible jobs and flexible jobs that are executed at their time of arrival.

Under the load shifting scenario, the optimised power and energy consumption are calculated using (\ref{eq:power_optimised}) and (\ref{eq:energy_optimised}). The formulation for the optimised power, $P^{\mathrm{IT}}_{\mathrm{opt}}(t)$, is piecewise. For the main 24 hour period, $t \in T$, the power consumption is calculated directly from the total optimised CPU utilisation, which includes the inflexible workload plus all flexible job tranches scheduled to run in that timeslot. For the 3 hour extension period, $t \in T^{\text{ext}} \setminus T$, the calculation is designed to isolate the impact of shifted workloads. To ensure a fair comparison over 24 hours against the base power demand, all of the workload originating within the 24 hour window is considered. Therefore, the formulation first calculates the total power draw in the extension slot and then subtracts the baseline power consumption $P^{\mathrm{IT}}_{\mathrm{base}}(t)$ for that same slot. This subtraction ensures that $P^{\mathrm{IT}}_{\mathrm{opt}}(t)$ during the extension window represents only the additional power demand attributable to jobs shifted from the original 24 hour period. Importantly, $P^{\mathrm{IT}}_{\mathrm{opt}}(t)$ can then be met by the grid or the UPS ESS.

In (\ref{eq:power_optimised}), $u(t\!-\!j,k,s)$ is evaluated at $s=t$  so that only IT workload executed at time $t$ contributes to power. The same $|_{s=t}$ convention is used wherever $u(t\!-\!j,k,s)$ appears, to account for execution at the time of interest.

IT jobs cannot be scheduled before their arrival or after their delay window, which is ensured by (\ref{eq:feasibility}). While performing IT workload shifting, it is essential to ensure that all jobs are executed and completed within their maximum delay tolerance. These conditions are satisfied through (\ref{eq:alpha_sum}) and (\ref{eq:job_completion}). Constraint (\ref{eq:alpha_sum}) ensures that the fractions $\alpha_{t,k}$ assigned to each tranche "k" of a job sum to one, distributing the entire job demand across its tranches. Constraint (\ref{eq:job_completion}) guarantees that the total computational work performed for each tranche over its execution window matches the required demand for that tranche.

Finally, constraint (\ref{eq:cpu_capacity_new}) enforces per–time-slot CPU capacity. At each slot $t$, the CPU utilisation equals the inflexible load plus the contributions of flexible tranches that originated earlier $(t-j)$ and are executed at $t$. This is expressed by evaluating $u(t\!-\!j,k,s)$ at $s=t$. The summed utilisation is bounded by $u_{\max}$, ensuring no schedule exceeds the available CPU capacity at the time of execution.

\begin{figure*}[!t]
\label{fig:it-equations}
\normalsize
\begin{flalign}
& P^{\mathrm{IT}}_{\mathrm{opt}}(t) = \begin{cases} P^{\mathrm{IT}}_{\mathrm{idle}} + (P^{\mathrm{IT}}_{\mathrm{max}} - P^{\mathrm{IT}}_{\mathrm{idle}}) \times \left( u^{\text{inflex}}(t) + \sum_{j=0}^{D_{\text{max}}} \sum_{k=1}^{K} u(t-j,k,s)\big|_{s=t} \right)^{1.32}, & \quad t \in T \\ P^{\mathrm{IT}}_{\mathrm{idle}} + (P^{\mathrm{IT}}_{\mathrm{max}} - P^{\mathrm{IT}}_{\mathrm{idle}}) \times \left( u^{\text{inflex}}(t) + \sum_{j=0}^{D_{\text{max}}} \sum_{k=1}^{K} u(t-j,k,s)\big|_{s=t} \right)^{1.32} - P^{\mathrm{IT}}_{\mathrm{base}}(t), & \quad t \in T^{\text{ext}} \setminus T \end{cases} \label{eq:power_optimised} & 
\end{flalign}
\end{figure*}

\begin{figure}[!t]
\centering
\label{fig:it-workload-equations}
\normalsize
\begin{align}
& 0 \le u(t,k,s) \le u_{\text{max}}, \quad \forall t \in T, \forall k \in K, \forall s \in W_{t,k} \label{eq:util_bounds} & \\[0.4\baselineskip]
& P^{\mathrm{IT}}_{\mathrm{base}}(t) = P^{\mathrm{IT}}_{\mathrm{idle}} + (P^{\mathrm{IT}}_{\mathrm{max}} - P^{\mathrm{IT}}_{\mathrm{idle}}) \times (u^{\text{inflex}}(t) + \nonumber \\ 
& \qquad u^{\text{flex,base}}(t))^{1.32}, \quad \forall t \in T^{\text{ext}} \label{eq:power_base} & \\[0.4\baselineskip]
& E^{\mathrm{IT}}_{\mathrm{base}}(t) = \Delta t \cdot P^{\mathrm{IT}}_{\mathrm{base}}(t), \quad \forall t \in T^{\text{ext}} \label{eq:energy_base} & \\
& E^{\mathrm{IT}}_{\mathrm{opt}}(t) = \Delta t \times P^{\mathrm{IT}}_{\mathrm{opt}}(t), \quad \forall t \in T^{\text{ext}} \label{eq:energy_optimised} & \\[0.4\baselineskip]
& u(t,k,s) = 0 \quad \text{if } s < t \text{ or } s > t + D_k, \nonumber \\ 
& \qquad \forall t \in T, \forall k \in K, \forall s \in T^{\text{ext}} \label{eq:feasibility} & \\
& \sum_{k=1}^{K} \alpha_{t,k} = 1, \quad \forall t \in T \label{eq:alpha_sum} & \\[0.4\baselineskip]
& \sum_{s \in W_{t,k}} u(t,k,s) \cdot \Delta t = R_t \cdot \alpha_{t,k}, \quad \forall t \in T, \forall k \in K \label{eq:job_completion} & \\
& u^{\text{inflex}}(t) + \sum_{k=1}^{K} \sum_{j=0}^{D_{k}} u(t-j, k, s)\big|_{s=t} \leq u_{\text{max}}, \quad \forall t \in T^{\text{ext}} \label{eq:cpu_capacity_new} &
\end{align}
\end{figure}

\subsection{UPS ESS Mathematical Model and Constraints}

 The key decision variables in this model are the state of charge ($E^{\mathrm{UPS}}(t)$), the charging and discharging powers ($P^{\mathrm{UPS}}_{\mathrm{ch}}(t)$, $P^{\mathrm{UPS}}_{\mathrm{disch}}(t)$), and their corresponding binary status indicators ($z^{\mathrm{UPS}}_{\mathrm{ch}}(t)$, $z^{\mathrm{UPS}}_{\mathrm{disch}}(t)$). The dynamic operation of a UPS ESS, transitioning from time \(t\) to \(t+\Delta t\), is governed by the set of equations and inequalities shown in (\ref{eq:15})--(\ref{eq:22}).

The energy level of the batteries at the end of time interval \(t\) is calculated based on the charging and discharging power applied during that interval, as formulated in Equation (\ref{eq:15}) \cite{Takci2024Chapter}.
The energy level of the batteries must be maintained within their predefined operational boundaries as shown in (\ref{eq:16}) \cite{Takci2024Chapter}. The SoC limits prevent overcharging and deep discharging, as illustrated in (\ref{eq:17}). Furthermore, to ensure energy balance over the optimisation period, the energy level at the start and end of the cycle are constrained to be equal, as specified in (\ref{eq:18}).

The charging and discharging powers are constrained within the allowable limits of the UPS ESS, as shown in (\ref{eq:19}) and (\ref{eq:20}). Equation (\ref{eq:21}) guarantees that charging and discharging do not occur simultaneously \cite{Takci2024Chapter}. Net Power, as shown in (\ref{eq:22}), is defined as the difference between charging and discharging power, and is a single, signed variable representing the total instantaneous power exchange in the UPS. This allows for a physically accurate constraint on the UPS converter, whether it is charging from the grid, discharging to the grid, or powering the IT equipment to reduce the grid load. \cite{Hashmi2024ArXiv}.

\begin{figure}[!t]
\centering
\label{fig:ups-equations}
\normalsize
\begin{flalign}
& E^{\mathrm{UPS}}(t+\Delta t) = E^{\mathrm{UPS}}(t) + (\eta^{\mathrm{UPS}}_{\mathrm{ch}} \cdot P^{\mathrm{UPS}}_{\mathrm{ch}}(t) \cdot \Delta t) - \notag\\ 
& \left(\frac{P^{\mathrm{UPS}}_{\mathrm{disch}}(t)}{\eta^{\mathrm{UPS}}_{\mathrm{disch}}} \cdot \Delta t\right), \quad \forall t \in T^{\text{ext}} \label{eq:15} \\[0.4\baselineskip]
& E^{\mathrm{UPS}}_{\mathrm{min}} \le E^{\mathrm{UPS}}(t) \le E^{\mathrm{UPS}}_{\mathrm{max}}, \quad \forall t \in T^{\text{ext}} \label{eq:16} \\[0.4\baselineskip]
& E^{\mathrm{UPS}}_{\mathrm{min}} = SoC_{\mathrm{min}} \cdot E^{\mathrm{UPS}}_{\mathrm{base}}, \quad E^{\mathrm{UPS}}_{\mathrm{max}} = SoC_{\mathrm{max}} \cdot E^{\mathrm{UPS}}_{\mathrm{base}} \label{eq:17} \\
& E^{\mathrm{UPS}}_{\mathrm{start}} = E^{\mathrm{UPS}}_{\mathrm{end}} = (50\%) \cdot E^{\mathrm{UPS}}_{\mathrm{base}} \label{eq:18} \\[0.4\baselineskip]
& z^{\mathrm{UPS}}_{\mathrm{ch}}(t) \cdot P^{\mathrm{UPS}}_{\mathrm{ch,min}} \leq P^{\mathrm{UPS}}_{\mathrm{ch}}(t) \le \notag \\
& \quad z^{\mathrm{UPS}}_{\mathrm{ch}}(t) \cdot P^{\mathrm{UPS}}_{\mathrm{ch,max}}, \quad \forall t \in T^{\text{ext}} \label{eq:19} \\[0.4\baselineskip]
& z^{\mathrm{UPS}}_{\mathrm{disch}}(t) \cdot P^{\mathrm{UPS}}_{\mathrm{disch,min}} \leq P^{\mathrm{UPS}}_{\mathrm{disch}}(t) \le \notag \\
& \quad z^{\mathrm{UPS}}_{\mathrm{disch}}(t) \cdot P^{\mathrm{UPS}}_{\mathrm{disch,max}}, \quad \forall t \in T^{\text{ext}} \label{eq:20} \\[0.4\baselineskip]
& z^{\mathrm{UPS}}_{\mathrm{ch}}(t) + z^{\mathrm{UPS}}_{\mathrm{disch}}(t) \le 1, \quad \forall t \in T^{\text{ext}} \label{eq:21} \\
& P^{\mathrm{UPS}}_{\mathrm{net}}(t) = P^{\mathrm{UPS}}_{\mathrm{ch}}(t) - P^{\mathrm{UPS}}_{\mathrm{disch}}(t),\quad \forall t \in T^{\text{ext}}\label{eq:22}
\end{flalign}
\end{figure}

This constraint model effectively captures the physical limitations of the power converter. If the battery is charging, then $P^{\mathrm{UPS}}_{\mathrm{disch}}(t)=0$ and $P^{\mathrm{UPS}}_{\mathrm{net}}(t)=P^{\mathrm{UPS}}_{\mathrm{ch}}(t)$. If it is discharging, then $P^{\mathrm{UPS}}_{\mathrm{ch}}(t)=0$ and $P^{\mathrm{UPS}}_{\mathrm{net}}(t)=-P^{\mathrm{UPS}}_{\mathrm{disch}}(t)$. If idle, both powers are zero.

\subsection{Cooling System Model} \label{sec: Cooling Infrastructure Flexibility} 
This section models DC cooling infrastructure to assess its flexibility potential. The methodology is based on the validated thermodynamic model by Cupelli et al. \cite{cupelli2018data}, which is adapted in this work by incorporating a Thermal Energy Storage (TES) tank and scaling the parameters for a 1 MW IT capacity data centre.
Flexibility is harnessed from two primary sources. The TES tank, which provides downward flexibility by charging (increasing electrical load) and upward flexibility by discharging (allowing the chiller to reduce its load). The second is the inherent thermal mass of the DC components, such as the IT servers, racks, and air in the DC. By allowing temperatures to fluctuate between $18$ and $22.5$ degrees, the masses act as a thermal buffer. The thermal buffer allows the chiller's power consumption to be modulated without affecting IT operations.
The cooling system of the DC is considered a closed-loop thermodynamic system where a central chiller provides cooled water to a CRAC unit, either directly or via the TES tank. This air is circulated through the DC to remove heat generated by the IT equipment, with a constant air mass flow rate, $\dot{m}_{a}$. The chiller is considered the main variable electrical load, while the power consumption of auxiliary components is treated as a constant overhead  $P^{\mathrm{Grid}}_{\mathrm{OD}}(t)$. The governing equations describing the system's thermal dynamics and operational constraints are presented below.

\begin{figure*}[!t]
\label{fig:cooling-equations}
\normalsize
\begin{flalign}
& Q_{CA}(t) + Q_{HA}(t) + Q_{R}(t) + Q_{ITm}(t) = Q_{out}(t) + Q^{\mathrm{IT}}(t) - Q_{\mathrm{cool}}(t), \quad \forall t \in T^{\text{ext}} & \label{eq:thermal_balance} \\
& Q_{\mathrm{cool}}(t) = Q^{\mathrm{Chil}\text{-}\mathrm{CRAC}}(t) + Q^{\mathrm{TES}\text{-}\mathrm{CRAC}}(t), \quad \forall t \in T^{\text{ext}} & \label{eq:cooling_power_rel} \\
& Q^{\mathrm{Chil}\text{-}\mathrm{CRAC}}(t) = P^{\mathrm{Chil}\text{-}\mathrm{CRAC}}(t) \cdot \mathrm{COP}_{\mathrm{chiller}}, \quad \forall t \in T^{\text{ext}} & \label{eq:CRAC_cop} \\
& Q^{\mathrm{Chil}\text{-}\mathrm{TES}}(t) = P^{\mathrm{Chil}\text{-}\mathrm{TES}}(t) \cdot \mathrm{COP}_{\mathrm{chiller}}, \quad \forall t \in T^{\text{ext}} & \label{eq:tes_charge_cop} \\
& P^{\mathrm{Chil}\text{-}\mathrm{CRAC}}(t) + P^{\mathrm{Chil}\text{-}\mathrm{TES}}(t) \le P^{\mathrm{Chiller}}_{\mathrm{max}}, \quad \forall t \in T^{\text{ext}} & \label{eq:chiller_max_power} \\
& E^{\mathrm{TES}}(t) = E^{\mathrm{TES}}(t-1) + \left[ \left( \eta^{\mathrm{TES}}_{\mathrm{ch}} \cdot Q^{\mathrm{Chil}\text{-}\mathrm{TES}}(t-1) \cdot \Delta t \right) - \left( \frac{Q^{\mathrm{TES}\text{-}\mathrm{CRAC}}(t-1)}{\eta^{\mathrm{TES}}_{\mathrm{dis}}} \cdot \Delta t \right) \right], \quad \forall t \in T^{\text{ext}} & \label{eq:tes_energy_update} \\
& T_{Ain}(t) = T_{HA}(t-1) - \frac{Q_{\mathrm{cool}}(t-1)}{\dot{m}_{a} \cdot c_{pa}}, \quad \forall t \in T^{\text{ext}} & \label{eq:air_inlet_temp} \\
& T_{IT}(t) = T_{IT}(t-1) + 3600 \cdot \Delta t \cdot \left(\frac{P^{\mathrm{IT}}(t-1) - G_{cv}\bigl(T_{IT}(t-1) - T_{R}(t-1)\bigr)}{C_{IT}}\right), \quad \forall t \in T^{\text{ext}} & \label{eq:ite_temp} \\
& T_{R}(t) = T_{R}(t-1) + 3600 \cdot \Delta t \cdot \left(\frac{\dot{m}_{a} \cdot \kappa \cdot c_{pa}\bigl(T_{CA}(t-1) - T_{R}(t-1)\bigr) + G_{cv}\bigl(T_{IT}(t-1) - T_{R}(t-1)\bigr)}{C_{R}}\right), \quad \forall t \in T^{\text{ext}} & \label{eq:rack_temp} \\
& T_{CA}(t) = T_{CA}(t-1) + 3600 \cdot\Delta t \cdot \left(\frac{\dot{m}_{a} \cdot \kappa \cdot c_{pa}\bigl(T_{Ain}(t-1) - T_{CA}(t-1)\bigr) - G_{cd}\bigl(T_{CA}(t-1) - T_{out})}{C_{CA}}\right), \quad \forall t \in T^{\text{ext}} & \label{eq:ca_temp} \\
& T_{HA}(t) = T_{HA}(t-1) + 3600 \cdot \Delta t \cdot \left(\frac{\dot{m}_{a} \cdot \kappa \cdot c_{pa}\bigl(T_{R}(t-1) - T_{HA}(t-1)\bigr)}{C_{HA}}\right), \quad \forall t \in T^{\text{ext}} & \label{eq:ha_temp} \\
& Q_{\mathrm{cool}}(t) \le \bigl(T_{HA}(t) - T_{CA,min}\bigr) \cdot \dot{m}_{a} \cdot c_{pa}, \quad \forall t \in T^{\text{ext}} & \label{eq:cooling_power_constraint} \\[0.3\baselineskip]
& 0 \le Q^{\mathrm{Chil}\text{-}\mathrm{TES}}(t) \le z^{\mathrm{Chil}\text{-}\mathrm{TES}}(t) \cdot Q^{\mathrm{Chil}\text{-}\mathrm{TES}}_{\mathrm{max}}, \quad \forall t \in T^{\text{ext}} & \label{eq:tes_ch_limit} \\[0.3\baselineskip]
& 0 \le Q^{\mathrm{TES}\text{-}\mathrm{CRAC}}(t) \le z^{\mathrm{TES}\text{-}\mathrm{CRAC}}(t) \cdot Q^{\mathrm{TES}\text{-}\mathrm{CRAC}}_{\mathrm{max}}, \quad \forall t \in T^{\text{ext}} & \label{eq:tes_dis_limit} \\[0.3\baselineskip]
& z^{\mathrm{Chil}\text{-}\mathrm{TES}}(t) + z^{\mathrm{TES}\text{-}\mathrm{CRAC}}(t) \le 1, \quad \forall t \in T^{\text{ext}} & \label{eq:tes_binary_logic} \\[0.3\baselineskip]
& E^{\mathrm{TES}}(|T|) = E^{\mathrm{TES}}(1) & \label{eq:tes_energy_cycle} \\[0.3\baselineskip]
& E^{\mathrm{TES}}(t) \le E^{\mathrm{TES}}_{\mathrm{max}}, \quad \forall t \in T^{\text{ext}}& \label{eq:tes_energy_max} \\[0.3\baselineskip]
& T_{j,min} \le T_{j}(t) \le T_{j,max}, \qquad \forall j \in \{\text{Ain, CA, R, IT, HA}\}, \quad \forall t \in T^{\text{ext}} & \label{eq:temp_constraints}
\end{flalign}
\hrulefill
\vspace*{4pt}
\end{figure*}


Equation \eqref{eq:thermal_balance} represents the overall thermal energy balance, where the rate of change of energy stored in the DC's thermal masses (aisles, racks, ITE) equals the net heat flow from the environment, ITE, and cooling system, where $Q_{\mathrm{out}}(t) = G_{cd} (T_{\mathrm{out}} - T_{CA}(t))$. It is assumed that all ITE electrical power converts to heat, i.e., $P^{\mathrm{IT}}(t) = Q^{\mathrm{IT}}(t)$. The discrete-time temperature dynamics for the supply air, ITE, racks, cold aisle, and hot aisle are governed by Equations \eqref{eq:air_inlet_temp} through \eqref{eq:ha_temp}, employing an explicit forward Euler method.

The primary distinction between the equation for cold aisle temperature (\ref{eq:ca_temp}) and hot aisle temperature (\ref{eq:ha_temp}), stems from the modelled hot aisle containment, which prevents any heat loss from the hot aisle to the external environment. Equation \eqref{eq:cooling_power_rel} shows the total cooling power is the sum of contributions from the CRAC and the discharging TES. Equations \eqref{eq:CRAC_cop} and \eqref{eq:tes_charge_cop} relate the chiller's electrical power consumption to the thermal cooling delivered, governed by its Coefficient of Performance ($\mathrm{COP}_{\mathrm{chiller}}$). The chiller's total power draw is capped by constraint \eqref{eq:chiller_max_power}. The cooling power is limited by \eqref{eq:cooling_power_constraint} to prevent overcooling the cold aisle. The TES state of charge is updated by Equation \eqref{eq:tes_energy_update}, accounting for charging/discharging efficiencies. The operational limits of the TES are defined in \eqref{eq:cooling_power_constraint} through \eqref{eq:tes_binary_logic}, which set maximum power levels and prevent simultaneous charging and discharging. The TES energy level is constrained by \eqref{eq:tes_energy_max}, while \eqref{eq:tes_energy_cycle} enforces a cyclic energy balance, ensuring the final state of charge equals the initial state.
To ensure safe operation, constraint \eqref{eq:temp_constraints} enforces the upper and lower temperature bounds for all thermal components.

\section{Case Studies: Integrated DC Model}\label{sec:integrated_model and case studies}

This section evaluates the integrated operation of a data centre  in which IT workload scheduling, UPS ESS dispatch, and cooling/TES control are co-optimised. Building on the component models in Section 3, a whole-facility formulation is assembled, linked by electric and thermal coupling constraints, and tested under realistic price signals. The scenarios are designed to answer two questions: (i) to what extent can an integrated optimum energy management algorithm reduce operating cost, and (ii) using this cost-optimised operation as a baseline, how much upward/downward flexibility can be offered to the power system without violating service levels. This two-step approach models a realistic scenario where an operator first determines their optimal 24 hour schedule and then uses any remaining capacity to participate in flexibility services to generate additional revenue.

A hypothetical data centre with 1 MW IT capacity and associated infrastructure is considered. The electricity price is defined by the day-ahead price profiles detailed in Table \ref{tb:energy_price}. The electricity price for the three hour extension period from 0–3 hours of the following day is the same as the electricity price between 0–3 hours given in Table \ref{tb:energy_price}. All parameter values used are detailed in Table \ref{tb:nomenclature}. 

\begin{table*}[!t]
\caption{Day-Ahead Energy Prices}
\label{tb:energy_price}
\centering
\resizebox{\textwidth}{!}{%
\small
\begin{tabular}{l|*{24}{c}}
\toprule
\textbf{Hour} & 0 & 1 & 2 & 3 & 4 & 5 & 6 & 7 & 8 & 9 & 10 & 11 & 12 & 13 & 14 & 15 & 16 & 17 & 18 & 19 & 20 & 21 & 22 & 23 \\
\midrule
\textbf{Price (GBP/MWh)} & 60 & 55 & 52 & 50 & 48 & 48 & 55 & 65 & 80 & 90 & 95 & 100 & 98 & 95 & 110 & 120 & 130 & 140 & 135 & 120 & 100 & 90 & 80 & 70 \\
\bottomrule
\end{tabular}
}
\end{table*}

\subsection{Scenario 1: Base Cost Calculation}\label{sec:scenario 1}
This scenario calculates the overall cost without any optimisation or flexibility utilisation. The purpose of this scenario is to define a base case that provides a benchmark for measuring the potential savings achieved through demand side flexibility. The total DC electricity cost is defined as:

\begin{flalign}
& \sum_{t \in T}
\bigl(
P^{\mathrm{Grid}}_{\mathrm{IT}}(t)
+ P^{\mathrm{Grid}}_{\mathrm{OD}}
+ P^{\mathrm{Chil}\text{-}\mathrm{CRAC}}(t)
\bigr) \cdot \Delta t \cdot \pi(t), && \\
& P^{\mathrm{Grid}}_{\mathrm{OD}} = E\left[P^{\mathrm{Grid}}_{\mathrm{IT}}(t) + P^{\mathrm{Chil}\text{-}\mathrm{CRAC}}(t)\right], \quad \forall t \in T &&
\end{flalign}

where $\pi(t)$ is the day-ahead spot energy price at time $(t)$. The UPS and TES tank are not utilised, since there is no optimisation in place to schedule their use, and so will not incur any cost in this scenario. $P^{\mathrm{Grid}}_{\mathrm{OD}}(t)$ is calculated as 7\% of the average base case power consumption. Additionally, all IT workload is considered inflexible and the temperature of the DC is constrained to a constant $22.5$ degrees. These adaptations result in a base case where the DC operates without utilising any flexibility.

\subsection{Scenario 2: Cost Minimisation}\label{sec:scenario 2}
This scenario calculates the cost-optimised operating conditions by shifting IT workloads over time, utilising the UPS as an energy storage system (ESS), and leveraging the thermal mass of the DC and TES tank for energy storage. This optimisation is highly dependent on the electricity pricing mechanism. In this study, a set of day-ahead spot electricity prices is utilised but further work could implement varying pricing strategies and evaluate the sensitivity of the results to different pricing models. The objective function is to minimise the DC total electricity cost, as defined in Equation (\ref{eq:cost_minimisation}). This is subject to some additional constraints presented in Equations (\ref{eq:43}-\ref{eq:46}). Equation (\ref{eq:43}) ensures that the power supplied to the IT equipment from the grid and the UPS sums to the optimised IT power demand ($P^{\mathrm{IT}}_{\mathrm{opt}}(t)$), which is defined in Equation (\ref{eq:power_optimised}). Power from auxiliary devices, $P^{\mathrm{Grid}}_{\mathrm{OD}}(t)$ is modelled as a $7\%$ of the average base case power consumption fraction. Equation (\ref{eq:46}) enforces the non-negativity of all power variables.

Note that the objective explicitly includes grid withdrawals (e.g. \(P^{\mathrm{Grid}}_{\mathrm{IT}}(t)\)) and charging powers (e.g. \(P^{\mathrm{UPS}}_{\mathrm{ch}}(t)\), \(P^{\mathrm{Chil}\text{-}\mathrm{TES}}(t)\)). UPS discharge \(P^{\mathrm{UPS}}_{\mathrm{disch}}(t)\) therefore does not appear directly in the cost sum; its economic effect is realised implicitly through Equation~(\ref{eq:43}), which reduces \(P^{\mathrm{Grid}}_{\mathrm{IT}}(t)\) when the UPS discharges.

\begin{figure*}[!t]
\normalsize
\begin{flalign}
& \min \sum_{t \in T^{\mathrm{ext}}}
\bigl(
P^{\mathrm{Grid}}_{\mathrm{IT}}(t)
+ P^{\mathrm{Grid}}_{\mathrm{OD}}(t)
+ P^{\mathrm{UPS}}_{\mathrm{ch}}(t)
+ P^{\mathrm{Chil}\text{-}\mathrm{CRAC}}(t)
+ P^{\mathrm{Chil}\text{-}\mathrm{TES}}(t)
\bigr) \cdot \Delta t \cdot \pi(t), & \label{eq:cost_minimisation} \\
& P^{\mathrm{IT}}_{\mathrm{opt}}(t) = P^{\mathrm{Grid}}_{\mathrm{IT}}(t) + P^{\mathrm{UPS}}_{\mathrm{disch}}(t), \quad \forall t \in T^{\text{ext}} & \label{eq:43} \\
& 0 \leq P^{\mathrm{Grid}}_{\mathrm{IT}}(t), P^{\mathrm{Grid}}_{\mathrm{OD}}(t), P^{\mathrm{UPS}}_{\mathrm{ch}}(t), P^{\mathrm{Chil}\text{-}\mathrm{CRAC}}(t), P^{\mathrm{Chil}\text{-}\mathrm{TES}}(t), \quad \forall t \in T^{\text{ext}} & \label{eq:46}
\end{flalign}
\hrulefill
\vspace*{4pt}
\end{figure*}

In this scenario, the integrated DC model is used in a MILP optimisation over a 24 hour time window and 3 hour extension period outlined in Section \ref{sec:Quantifying Maximum Temporal Flexibility from IT Workload Shifting}. To maintain the model's linearity, the non-linear function in Equation~(\ref{eq:power_optimised}) is linearized using a piecewise linear approximation with Special Ordered Sets of Type 2 (SOS2). This technique approximates the non-linear curve with a series of connected line segments. The model was implemented in Python using the Pyomo modelling library and solved using the SCIP optimisation solver. Each optimisation run was solved to optimality, with a typical solve time of approximately 5-10 seconds. The computations were performed on a machine equipped with an Intel Core Ultra 9 185H processor and 32.0 GB of RAM.

\subsection{Scenario 3: Flexibility Duration Calculation}\label{sec:scenario 3}
The third scenario evaluates the data centre's capacity to provide flexibility in response to a direct grid signal. This approach models a realistic situation where the grid operator requests a specific magnitude of power change for a certain duration, for example during planned balancing events or unforeseen emergencies. The optimised baseline from Scenario 2 is taken and from it the maximum duration for which the DC can sustain specified levels of flexibility magnitude is determined. This is calculated at any given time slot over the 24-hour period. Here, flexibility magnitude, $\Delta P$ (kW), is defined as the power deviation from the optimised baseline; with upward flexibility (load reduction) represented by negative magnitudes and downward flexibility (load increase) by positive magnitudes. For each combination of start time $t_0$ and requested flexibility magnitude $\Delta P$, the longest continuous period $\tau$ is determined over which the system can track a correspondingly modified grid-power trajectory while maintaining operational feasibility and respecting all physical and thermal constraints. This temporal measure of flexibility provides a key input for assessing the DC's potential participation in demand-response schemes and ancillary-services markets by quantifying its ability to meet specific grid needs.

The flexibility analysis takes the optimised baseline from Scenario \ref{sec:scenario 2} as the reference trajectory and fixes the initial states at the chosen start time $t_0$. The four delay tranches defined in Section \ref{sec:Quantifying Maximum Temporal Flexibility from IT Workload Shifting} are not sufficient to model the IT workload after it has been shifted. This is because each IT job has been shifted by a variable amount between $0$ time slots and the maximum delay tolerance allowed by the tranche. Therefore, each IT job has a new delay tolerance equal to its original delay tolerance minus the amount it has already been shifted. When considering IT workload across the original $4$ tranches, jobs can now have delay tolerances from $1$ - $12$ time slots. Therefore, the original tranche definitions can be discarded and replaced with a new definition which incorporates this change. The change in tranche and delay tolerance definitions is detailed in (\ref{eq:k_def}) and (\ref{eq:Dk_def}). 

The only other constraint change to the formulation described in Section~\ref{sec: methodology}, is the core requirement to deviate the power drawn from the grid, relative to the Scenario 2 baseline, at the requested level over the chosen window. Denoting the baseline grid power draw as $P^{\mathrm{Grid}}_{\mathrm{base}}(t)$ and the new grid power draw as $P^{\mathrm{Grid}}(t)$, the flexibility constraint is given in (\ref{eq:flex_constraint}).

\begin{flalign}
& k \in K = \{1, 2, 3, 4, 5, 6, 7, 8, 9, 10, 11, 12\} && \label{eq:k_def} \\
& D_k=\{1, 2, 3, 4, 5, 6, 7, 8, 9, 10, 11, 12\} && \label{eq:Dk_def} \\[0.3\baselineskip]
& P^{\mathrm{Grid}}(t) \begin{cases}
\begin{array}{l} 
\leq P^{\mathrm{Grid}}_{\mathrm{base}}(t) + \Delta P \pm P_{\mathrm{tol}},\quad \Delta P < 0,\\
\quad t_0 \leq t \leq t_0 + \tau + 12
\end{array} 
\\[0.6\baselineskip] 
\begin{array}{l} 
\geq P^{\mathrm{Grid}}_{\mathrm{base}}(t) + \Delta P \pm P_{\mathrm{tol}},\quad \Delta P > 0,\\ 
\quad t_0 \leq t \leq t_0 + \tau + 12
\end{array} 
\end{cases} && \label{eq:flex_constraint}
\end{flalign}

For each start time $t_0$ and flexibility magnitude $\Delta P$, the feasibility of providing the given $\Delta P$ is tested over a series of durations. The optimisation described in Scenario 2 is run for each duration. To do so, the optimisation window is clipped such that $t_0 \leq t \leq t_0 + \tau + 12$. The $+12$ is the maximum delay tolerance of any IT workload and corresponds to the recovery time window directly after the duration $\tau$. The flexibility provision is only deemed feasible if the DC can operate without breaching any constraints in the $12$ slot recovery window. All of the IT workload initially scheduled for completion in the recovery window is considered inflexible so that the DC can go back to the operating conditions defined in Scenario 2 after the recovery time. Additionally, the flexibility constraint is not applied in the recovery window so the grid power draw may vary relative to the optimised baseline.

If the result for a given duration $\tau$ at timeslot $t_0$ and flexibility magnitude $\Delta P$ is feasible, the DC can provide that level of flexibility for the given duration. The same feasibility test is then repeated for a longer duration. If the result is infeasible, the test is run for a shorter duration. The process is repeated until the maxima $\tau(t_0, \Delta P)$ is found. The duration is modified using binary-search, so that $\mathcal{O}(\log \tau)$ feasibility checks are required per scenario. This efficient approach minimises computational overhead. The assessment is repeated across a grid of start times spanning the operating day and across a set of upward and downward flexibility magnitudes.

Lastly, the cold aisle temperature range is extended from $18$--$22.5 ^\circ\mathrm{C}$ to $18$--$23^\circ\mathrm{C}$ in Scenario 3. The $0.5^\circ\mathrm{C}$ increase provides a thermal buffer zone and remains well within the $27^\circ\mathrm{C}$ upper limit proposed by ASHRAE \cite{tc2016data}.
\section{Results}

\subsection{Scenario 1 and 2 Results}
\begin{figure}[!t]
 \centering
 \subfloat[]{\includegraphics[width=\columnwidth]{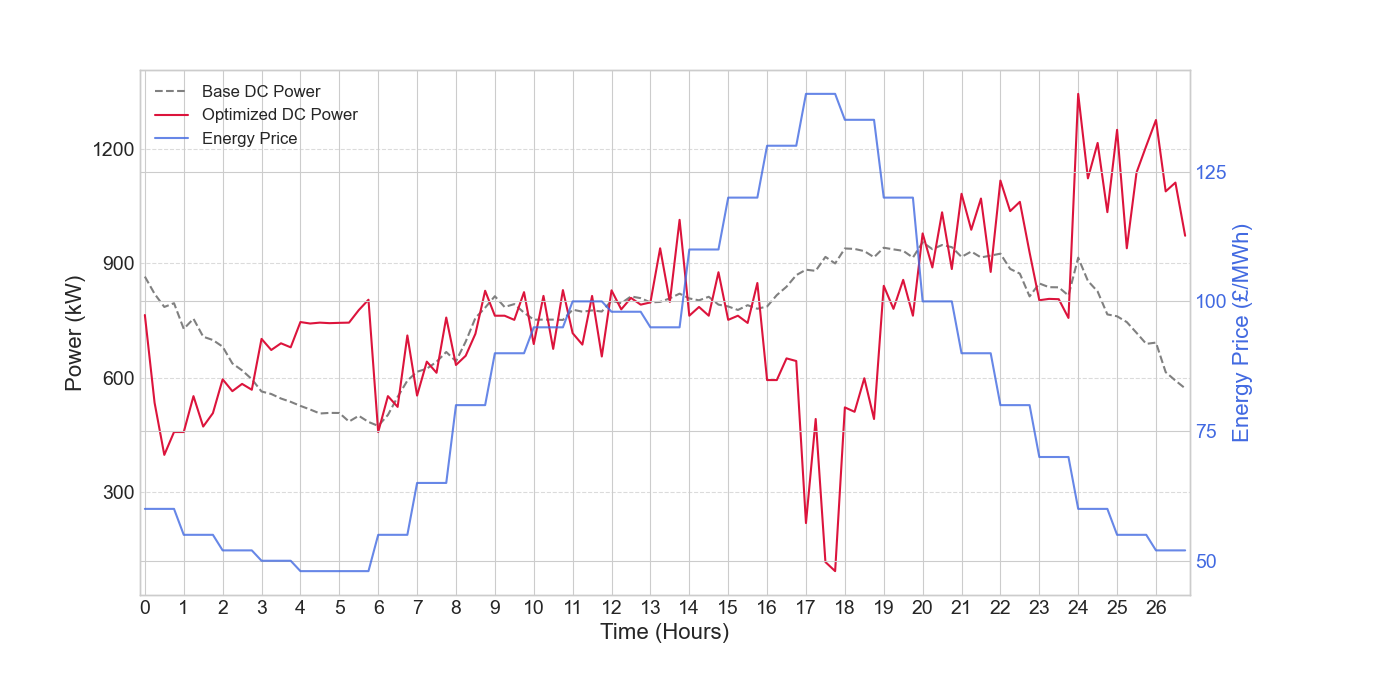}%
 \label{fig:power_chart}}
 \vspace{2mm} 
 \subfloat[]{\includegraphics[width=\columnwidth]{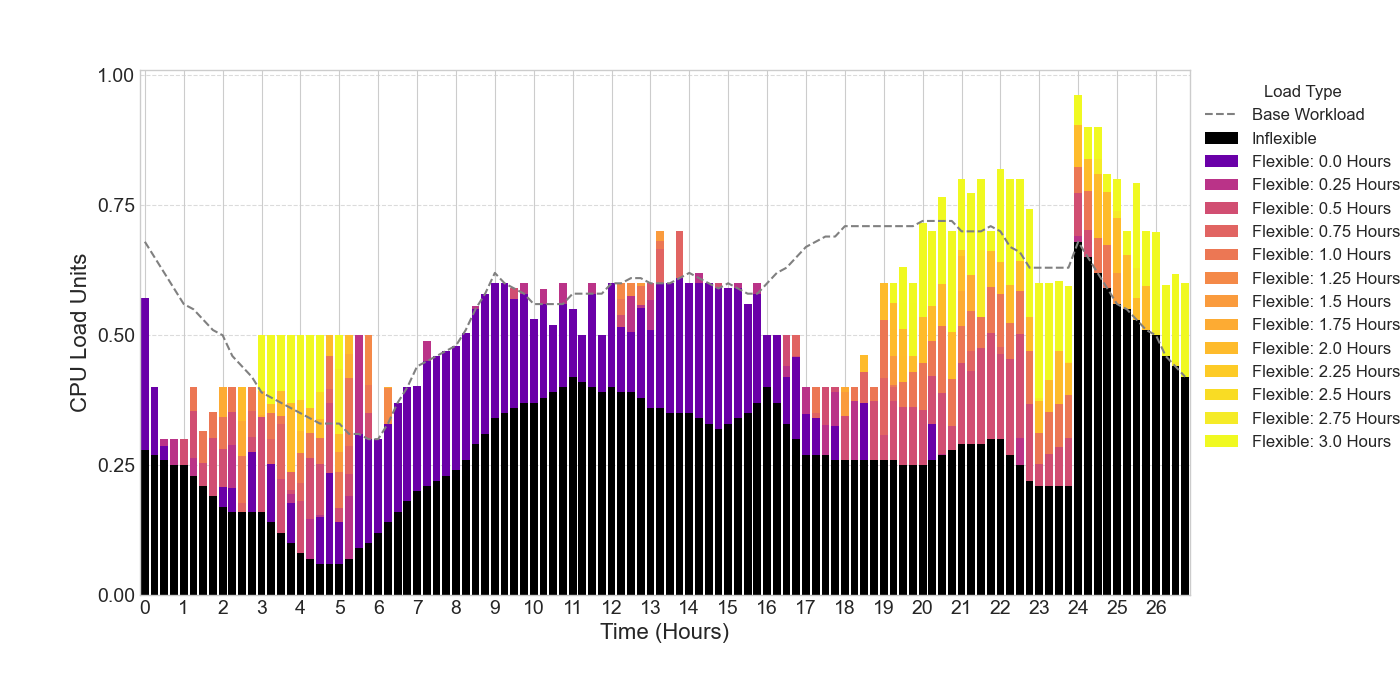}%
 \label{fig:it_load_chart}}
 \caption{(a) Power consumption profile and energy prices over the monitored period. (b) The corresponding IT workload distribution, shown as a stacked bar chart.}
 \label{fig:stacked_images}
\end{figure}

Figure \ref{fig:stacked_images} juxtaposes cost and workload profiles in subfigures (a) and (b) to highlight optimisation impacts. Subfigure (a) shows the base and optimised cost of operating the DC and the day-ahead electricity price in GBP over the simulation window. The total base and optimised costs are $1,659.54$ and $1,493.19$ GBP, respectively, representing a $ 166.34$ GBP cost saving. This cost saving is significant as it demonstrates that, with an appropriate management algorithm, existing data centre flexibility assets can be repurposed for cost minimisation, yielding savings of $10.02\%$ without additional investment, operational expenditure or sacrificing quality of service. Subfigure (b) shows the combined CPU utilisation of all the servers in the DC over time. The dotted grey line shows the CPU utilisation in the base case before IT workload shifting has taken place. The stacked bar chart shows the optimised CPU utilisation after IT workload shifting. The black bars show the inflexible workload which cannot be shifted. The coloured bars show the flexible workload that is either shifted between 15 minutes (labelled as Flexible: 0.25 hours) and 3 hours (labelled as Flexible: 3.0 Hours), or not shifted and so executed in its original timeslot. 

Subfigure (a) shows a considerable drop in operating costs between 16:30 and 20:30, which is a direct consequence of the optimisation algorithm shifting power consumption away from the peak energy price. Subfigure (b) shows how the IT workload originally scheduled in this time window has been shifted forward. The colours represent how far the workload has been shifted forward, with lighter colours showing workload that is shifted further. The y-axis shows the CPU utilisation of all the servers in the data centre, which has a maximum value of 1. The shifting shown in Figure \ref{fig:stacked_images}(b) considerably reduces power consumption from the IT workload at peak times, dramatically reducing DC operating costs.

The large amount of flexible load that is shifted by $3$ hours illustrates the optimiser delaying load for as long as possible to avoid higher energy prices. Equally, the long period between $06$:$00$ and $15$:$00$ of very minimal shifting is explained by an almost continuous rise in energy prices. IT workload deferral during a period of energy price increase only makes sense when the energy price drops soon after, as evidenced by the subsequent huge increase in load shifting as energy prices begin to fall. The coloured bars show continued IT workload shifting to the end of the extended time period to take advantage of the lower energy prices overnight. Additionally, at the beginning of the time window IT workload is shifted as energy prices fall during the early hours of the morning. 

\begin{figure}[!t]
 \centering
 \includegraphics[width=\columnwidth]{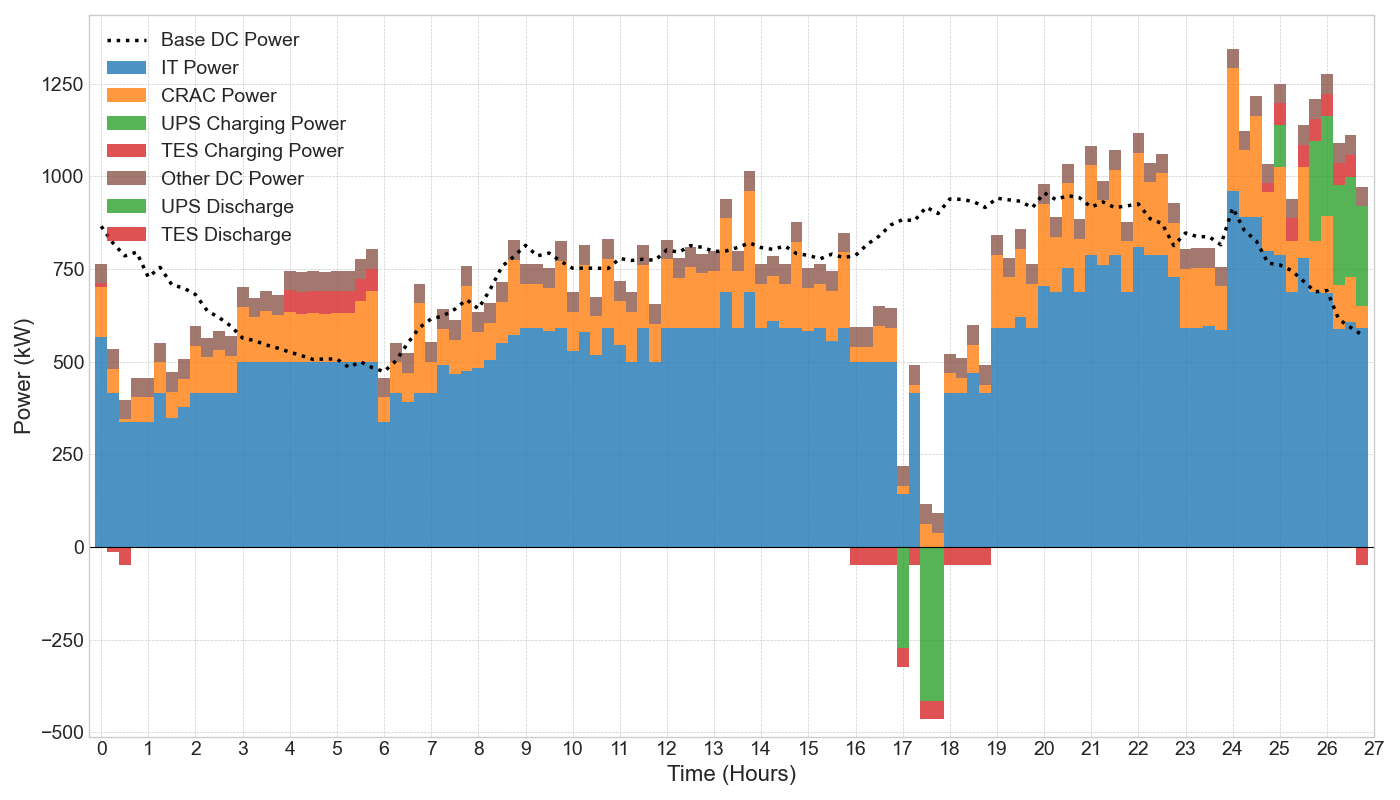}
 \caption{Optimised DC Component Power Consumption}
 \label{fig:power_resource_stack_chart}
\end{figure}

Figure~\ref{fig:power_resource_stack_chart} provides a detailed decomposition of the data centre's power consumption and resource dispatch strategy under the cost-minimisation objective of Scenario 2. The chart visualises the co-optimised management of various electrical loads and energy storage assets over the 24 hour period. The positive stacked areas represent the components of power drawn from the grid, including IT, CRAC, and charging power for the energy storage systems. The negative stacked areas indicate the power being discharged from the UPS and TES to serve internal loads, thereby reducing the need for grid electricity. The dotted black line represents the base DC power demand before optimisation, serving as a baseline to illustrate the impact of the flexibility measures.

Between 03:00 and 06:00 when electricity prices are low, the optimisation algorithm elects to draw more power from the grid than required in the base case. This surplus energy is partly used to charge the TES tank, effectively storing cheap energy for later use. Surplus power is also used to process additional IT workloads and consequently increase CRAC power consumption. Conversely, during the peak price period, around 16:00 to 19:00 hours, the grid power consumption is drastically reduced to a level far below the base demand. This reduction is achieved by dispatching the TES tank and UPS, and by shifting IT workload as shown in Figure \ref{fig:stacked_images}(b). The combined effect of these strategies is a huge decrease in DC operating costs around peak energy time and an overall cost saving of $>10\%$ over the full optimisation window. A data centre operator could use the proposed framework to optimise operating costs for the coming day and help stabilise the power system. 

\subsection{Scenario 3 Results}
\begin{figure*}[!t]
 \centering
 \includegraphics[width=\textwidth]{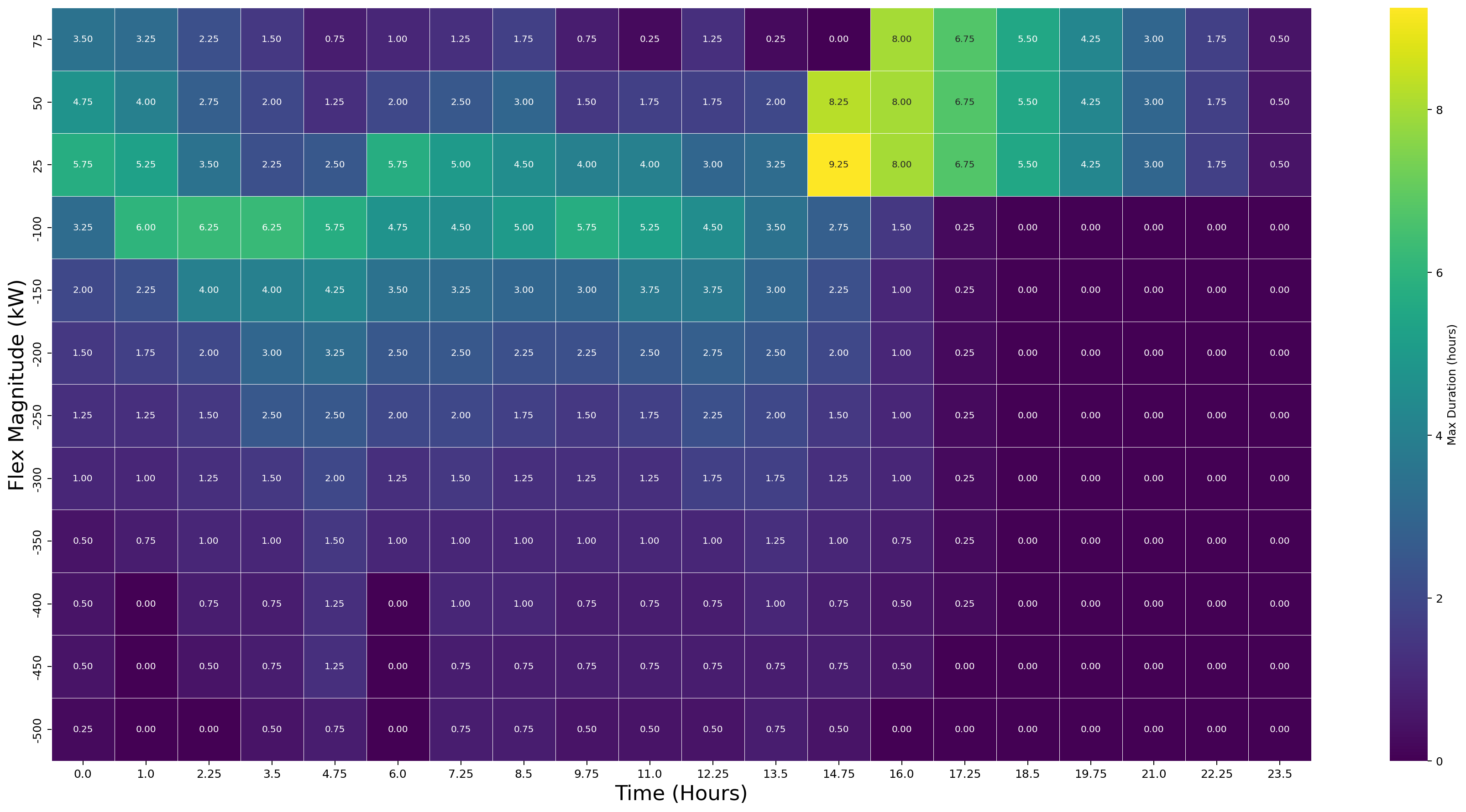}
 \caption{Flexibility Provision Magnitude and Duration}
 \label{fig:flex_duration_hourly_heatmap}
\end{figure*}

Scenario 3 proposes a comprehensive test to determine the maximum flexibility duration for a set of flexibility magnitudes and start times. The resulting data is three-dimensional and can be effectively visualised as a heatmap. This heatmap is shown in Figure \ref{fig:flex_duration_hourly_heatmap}, with $\tau$ represented by the colour and ($t_0$, $\Delta P$) shown on the (x, y) axes.

Figure~\ref{fig:flex_duration_hourly_heatmap} illustrates significant variations in flexibility duration $\tau$ with start time $t_0$ and deviation $\Delta P$. Peak $\tau$ values occur in the first 12 hours for small negative $\Delta P$ (upward flexibility) and in the afternoon for positive $\Delta P$ (downward flexibility). For positive $\Delta P$, durations remain low from midnight to 16:00 but surge thereafter.
This pattern stems from the optimised case shown in Figure~\ref{fig:stacked_images}, where operational costs (in GBP/15 min) plummet around 16:00 due to a sharp energy price spike, prompting the optimiser to minimise power consumption. This low power consumption enables a substantial increases in power draw during Scenario 3, thereby extending $\tau$ for positive $\Delta P$ starting at or after $t_0 = $16:00. The dramatic change in operating conditions in the optimised case around 16:00 is what gives Figure \ref{fig:flex_duration_hourly_heatmap} the distinctive change in $\tau$ values at this time. Another clear observation is that as the magnitude of the flexibility moves away from 0, the duration decreases. This is because the larger the magnitude the more flexibility potential is used up per timeslot, thereby reducing the value of $\tau$.

\begin{figure}[!t]
 \centering
 \includegraphics[width=\columnwidth]{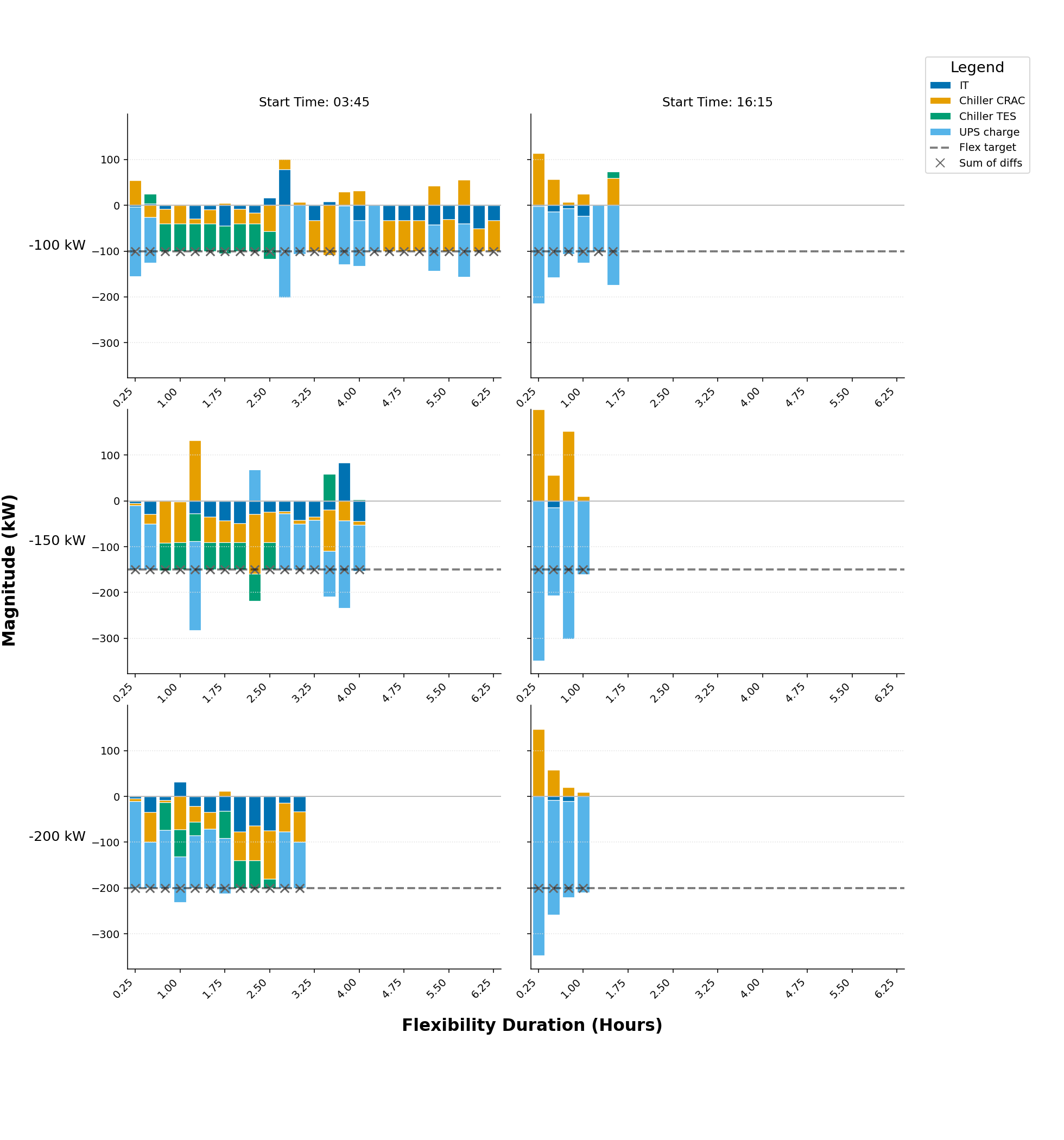}
 \caption{DC Component Contributions to Upward Flexibility}
 \label{fig:stacked_barcharts_neg}
\end{figure}

\begin{figure}[!t]
 \centering
 \includegraphics[width=\columnwidth]{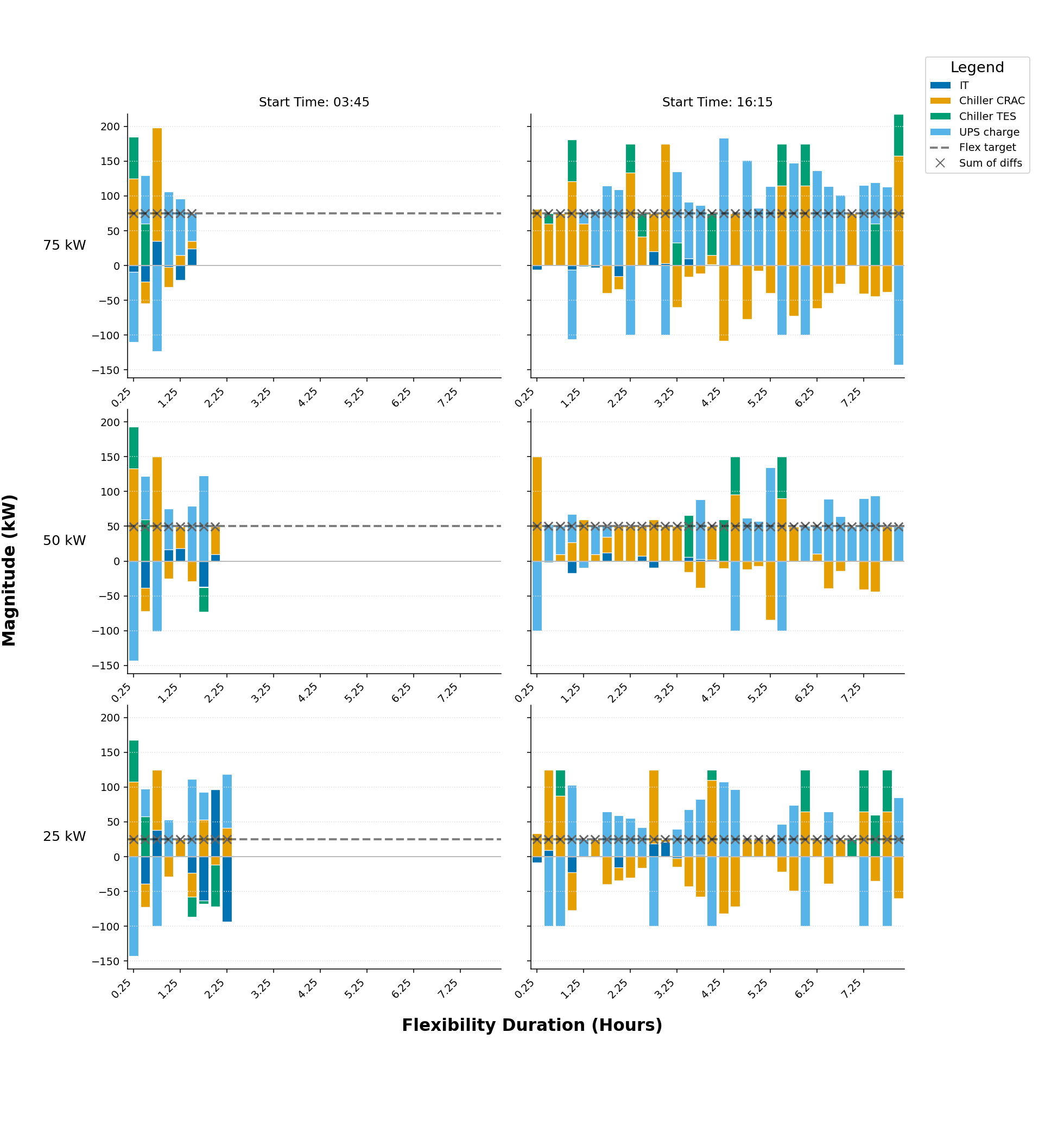}
 \caption{DC Component Contributions to Downward flexibility}
 \label{fig:stacked_barcharts_pos}
\end{figure}

Each of the elements in Figure \ref{fig:flex_duration_hourly_heatmap} can be decomposed and visualised as a stacked bar chart. A range of these stacked bar charts are shown in Figures \ref{fig:stacked_barcharts_neg} and \ref{fig:stacked_barcharts_pos}. In these plots, the duration is shown on the x-axis and the flexibility magnitude on the y-axis. Each stacked bar consists of the different sources of flexibility in the DC, illustrating how much flexibility is being provided by each DC asset throughout the flexibility provision. The dotted grey line shows the cumulative flexibility across all DC assets, which remains constant in each chart. Notably, in a considerable number of time slots, certain DC assets change their power consumption in the opposite direction to what is required. The other DC assets compensate by adjusting their power consumption to a greater extent, ensuring that the desired net flexibility magnitude is achieved. This is visible in the charts, where the stacked bars include both negative and positive components. This effect clearly illustrates the DC assets working in conjunction to achieve the desired net effect in the data centre.

Figure \ref{fig:stacked_barcharts_neg} shows six stacked bar charts arranged into a 3-by-2 grid. The rows and columns correspond to varying flexibility magnitudes and start times respectively. The colours show how much flexibility each DC asset is providing. Figure \ref{fig:stacked_barcharts_neg} shows negative $\Delta P$ (upward flexibility) and Figure \ref{fig:stacked_barcharts_pos} shows positive $\Delta P$ (downward flexibility). Figure \ref{fig:stacked_barcharts_neg} shows that $\tau$ depends strongly on $\Delta P$ and $t_0$. It is intuitive that the larger the flexibility magnitude the shorter the duration it can be provided for. $t_0$ has a large impact on duration due to the specific baseline from which the flexibility is being provided. The left-hand side of Figure \ref{fig:stacked_barcharts_neg} shows that the flexibility is provided by all DC components for varying durations and magnitudes. One notable result is that the IT workload and UPS tend to provide flexibility at different time slots. This corresponds to IT workload shifting providing flexibility, followed by the execution of those deferred IT jobs using the UPS to provide power such that the flexibility target can still be met. Additionally, cooling power reduction can also take place when there is IT workload shifting as less heat is generated by the servers. The TES tank, shown by the green bars, provides a large buffer for the required thermal power. This buffer enables flexibility to be provided across a range of DC operating conditions. In the Scenario 2 results which provide the baseline for this flexibility provision, the TES tank is being charged between 4:15 and 6:15 am. The green bars on the left-hand side of the figure in this time window show the TES tank providing flexibility by reducing the rate at which it is charging. 

Figure \ref{fig:stacked_barcharts_pos} shows the downward flexibility charts. As  with the upward flexibility charts, $\tau$ depends strongly on  $\Delta P$ and $t_0$. The CRAC, TES tank and UPS can all be seen to contribute a significant proportion of the downward flexibility. The IT workload makes up a much smaller proportion than in the upward flexibility case as it is not possible to reschedule workload to an earlier timeslot to increase its IT power consumption. One approach could be to include a `charge up' time before the flexibility window where earlier time slots could reschedule their workload to later time slots in the flexibility window. This is akin to the recovery time slot implemented after the flexibility provision. Another notable result is the large positive and negative magnitudes exhibited by different DC assets within the same timeslot. As mentioned previously, these assets provide opposing flexibility responses that cancel each other out to achieve the desired effect. This allows one asset to provide flexibility while the other ``recharges'' its flexibility potential, thereby demonstrating the value of an integrated DC model.

\section{Conclusion}

This paper presents an integrated, whole-facility framework that transforms a DC from purely a power consumer to a power prosumer that delivers a quantifiable flexibility provision. By co-optimising IT workload scheduling, UPS-ESS dispatch, and thermodynamic cooling with thermal energy storage, a cost-minimising day-ahead baseline was first established, achieving a 10.02\% reduction in operating costs. Building on this baseline, our core contribution is a duration-aware flexibility assessment that, for any start time ($t_0$) and requested power deviation ($\Delta P$), efficiently computes the maximum feasible duration ($\tau$) of flexibility provision while enforcing operational constraints and guaranteeing recovery. The resulting flexibility duration envelope reveals a strong temporal structure and a notable asymmetry. Upward flexibility (load reduction) is driven by deferring IT workload, allowing for a reduction in cooling power. The UPS manages the subsequent load recovery and the TES tank is modulated to provide additional flexibility. In contrast, downward flexibility (load increase) depends more heavily on increasing CRAC power consumption, supported by the TES buffer, and dispatching the UPS battery, as advancing IT workloads is inherently constrained without a pre-conditioning period. This framework successfully quantifies how much power a DC can shift and for precisely how long, turning abstract potential into the duration-certified flexibility required by market products. Figure \ref{fig:flex_duration_hourly_heatmap} provides a comprehensive set of results for the flexibility potential, detailing how the duration of flexibility provision changes depending on the magnitude required and specific start time. Results show a significant change in flexibility potential, with $100$ kW of upward flexibility possible for $6.8$ hours at $00$:$15$ and $0.2$ hours at $17$:$30$. This change reflects the varying operating conditions of the baseline from which flexibility is being provided. 

Promising future work includes incorporating explicit revenue stacking, exploring a ``pre-conditioning'' window to enable more symmetric IT participation in downward events, and aggregating facilities to a portfolio scale. Ultimately, the methodology presented offers a practical pathway for integrating data centres into power system operations, providing verifiable grid services while simultaneously reducing their own costs.

\bibliographystyle{IEEEtran}
\bibliography{references}

\section*{Acknowledgements}

This work was supported by the Engineering and Physical Sciences Research Council (EPSRC) and the Economic and Social Research Council (ESRC) through funding provided to the Energy Demand Research Centre (grant number EP/Y010078/1).

\section*{CRediT Authorship Contribution Statement}

\noindent \textbf{Mehmet Türker TAKCI:} Writing – original draft, Investigation, Methodology, Conceptualization.

\medskip
\noindent \textbf{James Day} Writing – original draft, Investigation, Methodology, Conceptualization.

\medskip
\noindent \textbf{Meysam Qadrdan:} Review \& editing, Supervision, Conceptualization.

\end{document}